\documentclass[]{aa}  

\usepackage{amsmath}
\usepackage{multirow}
\usepackage{tikz}
\usepackage{soul}
\usepackage{txfonts}
\usepackage{hyperref}
\usepackage{array}
\newcolumntype{P}[1]{>{\centering\arraybackslash}p{#1}}
\usepackage{graphicx}
\usepackage{xcolor}
\usepackage{natbib}
\usepackage{amssymb}
\usepackage[export]{adjustbox}
\usepackage{comment}

\pdfoutput=1
\newcommand{\HI}{{\ion{H}{i}}}
\newcommand{\red}[1]{\textcolor{black}{#1}}
\newcommand{\refcom}[1]{\textcolor{black}{#1}}
\newcommand{\lancom}[1]{\textcolor{black}{#1}}

\begin{document}

\title{Imaging the disk-halo interface of NGC 891: A 2.7\,kpc-thick molecular gas disk}
\subtitle{}
\author{D. Jiménez-López, S. García-Burillo, M. Querejeta, A. Usero, P. Tarrío}
\institute{Observatorio Astronómico Nacional, Madrid, Spain\\
              \email{d.jimenez@oan.es, s.gburillo@oan.es}
             }
\date{}
 
\abstract{
Halos surrounding spiral galaxies act as the bridges connecting the galactic disk and the intergalactic medium (IGM). They host a significant fraction of the baryonic mass in the Universe, and feedback from star formation (SF) or active galactic nuclei (\lancom{AGNs}) likely plays an important role in regulating this vertical baryonic component. Despite its importance, the contribution of extraplanar molecular gas remains poorly understood.}{
We aim to characterize the vertical extent and the kinematics  of molecular gas traced by CO(2-1) emission in the nearby ($D$ = 9.5~Mpc) spiral galaxy NGC\,891, one of the best\lancom{-}studied edge-on galaxies. 
We also compare our results with the extraplanar distribution of other tracers of baryonic matter, including atomic gas ($\HI$), \lancom{H$\alpha$-traced diffuse ionized gas (DIG)}, and dust maps from the literature.
}{
Our analysis is based on new CO(2–1) observations of NGC 891 obtained with the IRAM\,30m telescope. We mapped two 6 kpc $\times$ 6 kpc regions in the northeastern side and the area surrounding the galactic center. 
We applied a careful method to estimate and remove the residual contribution of the error beam to the CO cube.
} {
The vertical extent of the molecular gas is best described by a two-component Gaussian fit, consisting of a bright thin disk component \lancom{with a deconvolved full width at half maximum (FWHM) of $\leq$360\,pc} and a fainter thick disk component \lancom{with a deconvolved FWHM of $\simeq$ 1.1\,kpc}. Statistically significant ($>$3$\sigma$) CO(2-1) emission is detected 
up to 1.3-1.4\,kpc above the disk midplane.  We estimate that the thick molecular disk component contains up to $\sim$27\% of the total molecular gas mass of the galaxy.
The vertical extent of H$\alpha$ emission is similar to that of the molecular gas, whereas $\HI$ extends much further. 
}
{Our results demonstrate that SF-driven feedback in a non-starburst galaxy can lift significant amounts of molecular gas to large vertical distances. 
We interpret the presence of extraplanar molecular gas in NGC~891 in the framework of a galactic fountain scenario, in which material is expelled from star-forming regions and transported toward the outer halo. This is supported by optical images where dust columns rise from the inner disk, showing spatial correlation between CO (2-1), H$\alpha$, and dust.
}

\keywords{galaxies: individual: NGC 891 -- galaxies: ISM -- radio lines: galaxies -- galaxies: halos -- galaxies: structure -- galaxies: spiral}

\titlerunning{Disk-halo interface of NGC 891}

\maketitle

\nolinenumbers
\section{Introduction}
Galaxy halos surrounding spiral galaxies connect the inner galactic disk and the circumgalactic medium (CGM). This special environment is regulated by the balance between \lancom{the} inflow and outflow of matter and energy between these two systems.
Material found in the galactic halo may come from the CGM via cooling or be ejected from the galaxy midplane by, \lancom{for example}, supernova \lancom{(SN)} events and stellar clusters.
In-depth studies of these feedback mechanisms that regulate the cycle of extraplanar baryonic matter are essential to understanding the disk-halo interaction in spiral galaxies.
Ultimately, the disk-halo interaction is responsible for regulating the cycle of baryonic material and has important consequences on the star formation rate \lancom{(SFR)} and metallicity evolution of the galaxy.

Galaxy halos, along with the CGM, host a significant fraction of the baryonic mass of the Universe \citep{SL06, Li2018}. Previous studies suggest that the baryonic matter found in these environments may originate from outflows driven from the inner galactic disk. However, the dominant mechanisms are still uncertain. Some authors suggest that hot ionized gas is being pushed via stellar winds and \lancom{SN} explosions \citep{SF76, Bolatto13, Roy16}. Different works try to connect star formation (SF) with the distribution of hot gas \citep{Tullman06}, diffuse ionized gas (DIG, \citealt{Rossa03_a, Rossa03_b, Kamp06})\lancom{,} or dust \citep{HS00}.
Strong outflows from active galactic nuclei (\lancom{AGNs}) or SF can stir the interstellar medium \lancom{(ISM)} and easily produce vertically extended components (e.g., M82, \citealt{Walter_2002}; NGC\,1068, \citealt{sgb14}; Mrk\,231, \citealt{Aalto12, Aalto15}; Arp\,220, \citealt{BM18}). This is in principle more challenging for normal, \red{local (z=0),} Milky Way-like galaxies, where the \lancom{SF} is less intense. The processes regulating the ejection of material in those less active galaxies may take the form of galactic fountains \citep{Bregman80, Mel08, Armillotta_2016, fraternali06, Li23}, 
whereby gas is pushed up by stellar activity, travels through the halo\lancom{,} and eventually falls back to the disk. 
\red{This mechanism is more effective in dwarf galaxies, where the shallower potential makes the ejection of material via SF or SN activity \lancom{easier}.}
Recent works, such as the multiphase study of NGC\,4383 by \citet{cortese26}, provide support for galactic fountains as an important mechanism distributing material within the halo and regulating future \lancom{SF}.

In recent decades, most of our knowledge about the vertical structure of gas in nearby galaxies has come from observations of the ionized component traced by H$\alpha$ \citep{voig13, elliott26}, X-rays \citep{locatelly24}\lancom{,} and \lancom{the} radio continuum \citep{Krause18}, as well as the neutral $\HI$ gas \red{\citep{halogas_2011, Marasco19}} and dust \citep{Verst13}. 
Among the different phases, the molecular gas component remains the least studied, despite its crucial role in determining the total molecular hydrogen content and 
indirectly controlling
\lancom{SFRs}. 
Nearby edge-on spiral galaxies are the perfect laboratory to study the cycle of matter from the disk outward to the intergalactic medium (IGM). The unique orientation of highly inclined galaxies makes it easier to unambiguously determine the distribution of their out-of-plane gas thanks to 
the significant reduction of the dust extinction effects at high vertical distances 
and the reduction of ambiguities along the line of sight. For these reasons, many authors have chosen \red{galaxies of this kind} to study and characterize the different (out \lancom{and} in)flow processes and measure the vertical extent of the gas and dust. Galaxies such as NGC\,5775 \citep{reach2025}, NGC 2683 \citep{jiao25}, NGC 4244 \citep{zsch11}\lancom{,} and M82 \citep{wang24} have served as case studies \lancom{for} the outflow mechanisms and disk-halo dynamics in edge-on galaxies.

The nearby spiral galaxy NGC 891 (D $\simeq$ 9.5\,Mpc, \citealt{sancisiallen79}) is an excellent target for studying the disk-halo connection due to its nearly perfect edge-on inclination of $\sim$89$^\circ$  \citep{Xilouris_1998}.
It is similar to the Milky Way in total mass ($M= 1.4 \times10^{11}$\,M$_\odot$, including dark matter; \citealt{Oosterloo07}) and rotation speed \citep[$v_{\mathrm{rot}}^{\mathrm{max}}\simeq$ 230\,km\,s$^{-1}$;][]{Swaters_1997}, with \lancom{a} slightly higher SFR = 3.8 M$_\odot$yr$^{-1}$ \citep{halogas_2011, Evans_2022}.
\cite{Garcia-Burillo92} 
observed the CO(2-1) emission in several positions of the NGC\,891 halo with the Institut de \lancom{Radioastronomie Millimétrique (IRAM)} 30m telescope. They detected significant emission up to 1.3-1.6 kpc off the galaxy plane. This represented the first detection of a thick molecular disk in a non-starburst, non-active galaxy. The distribution and kinematics of the molecular gas were found to be consistent with a two-component structure consisting of a thin molecular disk 100-300 pc wide and a smoother, 2-3 kpc wide extended thick disk. The CO(1--0) map obtained with
the Nobeyama 40m antenna also shows the existence of a thick molecular gas disk extending up to $\simeq 1$~kpc \citep{Sof93}. NGC~891 also hosts a spatially extended $\HI$ halo that reaches up to 14\,kpc above of the galactic midplane. In the northeastern quadrant\lancom{,} it stretches even farther, forming a filament that extends to about 22\,kpc \citep{Oosterloo07}. This halo hosts roughly $30\%$ of the 
$\HI$ mass of the galaxy and shows many filamentary structures.

To study the galactic fountain process, it is necessary to examine the relationship between the molecular gas component and the DIG. It is thought that outflows originating from high star-forming regions are the engines that eject matter outward from the midplane. Many authors have studied this tracer to characterize the warm ionized medium of NGC\,891 (e.g.\lancom{,} \citealt{Rand97} with the William Herschel Telescope, WHT), and the vertical extent and kinematics of the DIG (e.g.\lancom{,} \red{\citealt{Kamphuis07, Kamp06}} with the TAURUS II imaging Fabry-Pérot spectrograph). \citet{HS00} also studied the DIG in NGC 891 and its relation with the dust distribution in the galaxy. The recent work of \citet{jwst24} confirmed that dust filaments connect to regions of high \lancom{SFRs} in the midplane of the galaxy, lending weight to the idea that galactic winds play a key role in the cycle of baryonic matter.

The emission from the  thick disk component of NGC~891 found by  \citet{Garcia-Burillo92} 
is significantly fainter than that of the thin disk. In order to reveal similar components in other galaxies, high-sensitivity observations would be required. The \lancom{full width at half maximum (FWHM)} thickness values of the molecular gas in the Milky Way reach only $\sim$50-160\,pc, increasing with galactocentric radius over 0–11 kpc \citep{Dame01, Sofue06, Mertsch21, Soding25}.  However, the existence of a thick molecular component in our Galaxy similar to the one found in NGC~891 might have been missed by CO surveys. \citet{Patra20} suggested the presence of a thin+thick gas disk structure in spiral galaxies, although the authors noted that limitations in spatial resolution and the galaxy inclination constraints made it difficult to confirm this scenario. Furthermore, \citet{pety13} also suggested the existence of a diffuse molecular gas phase as a thick disk that extends hundreds of parsecs outside the galactic plane of M~51, 
but using indirect methods due to the close to face-on orientation of this galaxy. Other studies also pointed to the existence of thick molecular disks to explain the kinematics of molecular gas in different samples of nearby spiral galaxies (e.g.\lancom{,} \citealt{Becq1997, CP13, CP15, CP16}).  

This work represents a follow-up project of the study of \citet{Garcia-Burillo92}. Unlike this earlier work, which relied on a limited number of vertical cuts 
perpendicular to the major axis to probe the vertical extent of molecular gas, our new dataset provides full coverage of two 2.2 $\arcmin$ $\times$ 2.2 $\arcmin$ (6\,kpc $\times$ 6\,kpc) fields around the central and northeastern regions of the galaxy. The enhanced sensitivity of the HEterodyne Receiver Array (HERA) used in our work significantly improves the data quality and provides a more robust framework for studying the vertical extent of molecular gas in NGC~891. Furthermore, the measurements of the error beam of \citet{Kramer13}, which are used in this work to estimate its potential contribution to the observed signal,  were mostly simultaneous 
with the CO observations. Moreover, the fact that we are fully sampling the targeted region ensures a better characterization of the error beam contribution across most of the disk compared to the previous estimates of \citet{Garcia-Burillo92}.

The paper is structured as follows. Section~\ref{Data_sample} describes the IRAM 30m observations and the ancillary data used in this work. The formalism used to characterize the error beam contribution to the CO flux, as well as  the derivation of the uncontaminated map used in this work\lancom{,} are described in Section~\ref{Error-beam}.
Section~\ref{res_disc} presents the main results, including a description of the distribution and kinematics of the extraplanar molecular gas. 
\lancom{In addition, the CO maps are compared with images of other ISM tracers, including the $\HI$, H$\alpha$ and dust distribution in the galaxy.}
Section~\ref{Discussion} discusses the implications for the origin of the extraplanar emission and the processes shaping its vertical structure. Section~\ref{Conclusions} summarizes our main conclusions.

\section{Observations}\label{Data_sample}

\subsection{CO (2-1) observations}
\begin{table*}[t]
\tiny
\caption{Setup used in the IRAM 30m telescope for the CO(2–1) observations of NGC 891.}
\centering
\begin{tabular}{ |P{1.55cm}||P{1.2cm}|P{1cm}|P{.8cm}|P{1cm}|P{1.4cm}|P{1.1cm}|P{1.2cm}|P{1cm}|P{.8cm}|  }
 \hline
  polarization & frequency (GHz) & $T^*_A$ (mK) & \lancom{RMS} (mK) & $\Delta v$ (km$\cdot$s$^{-1}$) & map size ($\Delta x$ $\times$ $\Delta y$) & mapping mode & switching mode & pwv (mm) & T (hours) \\
 \hline
 HERA 1+2 & 230.128 & 11-600 & 3.5 & 20 & 2.2' $\times$ 2.2' & OTF & PSw & 2 & 80h \\
 \hline
\end{tabular}

    \label{tab:setup_data}
\end{table*}

\begin{table*}[t]
\centering
\caption{\red{Error beam parameters.}}
\begin{tabular}{ 
 |l||c|c|c|c|  }
 \hline
   & Main beam & Error beam \#1 & Error beam \#2 & Error beam \#3 \\
 \hline
 Beam width $\theta$ & 10 \arcsec & 56 \arcsec & 217 \arcsec & 759 \arcsec \\
 Integrated relative power P [\%] & 69 & 5 & 13 & 13 \\
 \hline
\end{tabular}
\tablebib{\citet{Kramer13}}
    \label{tab:errorbeam_params}
\end{table*}

The present study is based on observations carried out with the IRAM 30m telescope using the HERA receiver array during two observing sessions, spanning 2008-2010. We covered two adjacent 2.2\,\arcmin $\times$ 2.2\,\arcmin (6\,kpc $\times$ 6\,kpc) fields of the galaxy using similar instrumental setups (see Table~\ref{tab:setup_data}). The first observing run targeted the northeastern half of the galaxy, while the second observing run focused on the region surrounding the galactic center. Either proposal was allocated 40\,h of observing time each. Based on previous CO(2-1) observations of NGC\,891 which detected out-of-plane emission with \lancom{a signal-to-noise ratio (S/N) of} $\sim$ 3, we aimed for a \lancom{S/N}$\sim$ 5 across the expected width of the thick molecular disk. Our observations, with a beam size of 11.2$\arcsec \times$ 11.2$\arcsec$ (516\,pc$\times$516\,pc), achieved an \lancom{root mean square (RMS) noise} of 3.5\,mK ($T^*_A$) in $\Delta v=20$\,km~s$^{-1}$-wide channels. 
Although the \lancom{RMS} noise was not entirely uniform across the field, we adopted an average value representative of the northern and southern halves. 
During the data reduction, first-order baselines were fit and subtracted to the spectra by defining a velocity window encompassing the emission from the midplane and excluding it from the fit.
The resulting data cube was re-gridded to a pixel size of $2\arcsec$ along both spatial axes. All images of the galaxy presented in this work show maps that have been rotated to align the galaxy's major axis \lancom{projected angle} \citep[\red{\lancom{PA}$\simeq203^\circ$}; e.g.,][]{Garcia-Burillo92, Scoville93} with the adopted $x$ axis.  We used the coordinates of the dynamical center of the galaxy for the 
rotation: $\alpha_{\rm 2000}=02^{\rm h}22^{\rm m}33^{\rm s}.063$, $\delta_{\rm 2000}=42^{\circ}20' 52".91$ \citep{Garcia-Burillo92, Scoville93}. Vertical offsets relative to the midplane are measured along the adopted $z$ axis.

\subsection{Ancillary data}

This work benefits from previous studies of other ISM components in NGC\,891, in particular the neutral hydrogen ($\HI$),  the ionized gas (H$\alpha$)\lancom{,} and the dust. Together, these tracers provide a comprehensive \lancom{multiphase} framework against which we compare the distribution and kinematics of molecular gas presented in this paper.

\subsubsection{WSRT $\HI$ observations}

The neutral hydrogen data used in this work come from the deep observations of NGC\,891 presented by \citet{Oosterloo07}. These observations are part of the \lancom{Hydrogen Accretion in LOcal GAlaxieS (HALOGAS)} survey \citep{halogas_2011} and were obtained with the Westerbork Synthesis Radio Telescope (WSRT). They rank among the deepest $\HI$ observations of an external galaxy to date.

The observations consisted of 20$\times$12~h integrations in multiple array configurations, covering a bandwidth of 10~MHz over 1024 spectral channels. The final processed high-resolution cube has a spatial resolution of 22.4$\arcsec$$\times$16.0$\arcsec$, an \lancom{RMS} noise per channel of 0.090~mJy~beam$^{-1}$, and a minimum detectable column density (3$\sigma$) of $1.3\times10^{19}$~cm$^{-2}$
\refcom{over a $\Delta v$ = 16.4 km s$^{-1}$ velocity resolution}.

$\HI$ emission is detected up to vertical distances of $\simeq22$~kpc above the disk midplane. The $\HI$ mass in the extended component represents $\sim30\%$ of the galaxy’s total $\HI$ mass.

\subsubsection{WIYN H$\alpha$ and BVI imaging}

To trace the warm ionized medium and its morphology above the NGC\,891 disk, we use the deep H$\alpha$ and \emph{BVI} broadband images presented by \citet{HS00}, obtained with the \lancom{Wisconsin-Indiana-Yale-NOIRLab (WIYN)} 3.5~m telescope. These sub-arcsecond resolution images reveal detailed diffuse and filamentary H$\alpha$ emission and a network of dust filaments extending up to $\simeq2$~kpc from the midplane. 

\subsubsection{Fabry–Pérot H$\alpha$ spectroscopy}

Finally, we incorporate H$\alpha$ Fabry–Pérot observations from \citet{Kamp06} to obtain kinematic information on the warm ionized medium and to compare the vertical extent of the H$\alpha$ halo with that of the molecular gas. The data were obtained with the TAURUS II Fabry–Pérot interferometer on the 4.2~m William Herschel Telescope (WHT) at La Palma.
These observations provide angular and spectral resolution sufficient to resolve the velocity field of the ionized gas across the disk–halo interface. 


\begin{figure*}[htbp]
\centering
\includegraphics[width=1\linewidth]{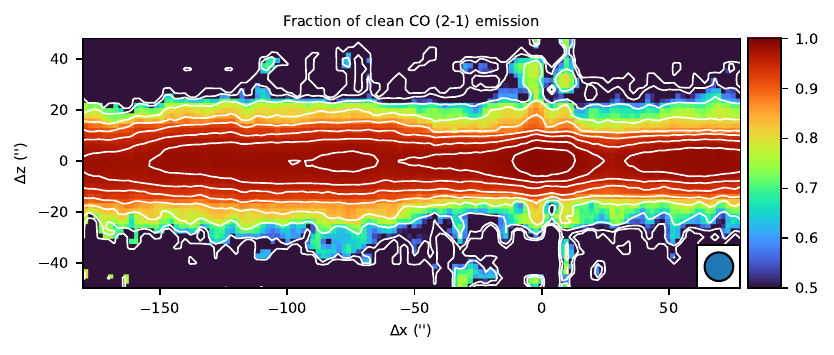}
\caption{\label{fig:contamination} Comparison of the observed CO\lancom{-}integrated intensity map (in white contours: from $3\sigma$ to 600$\sigma$, logarithmically spaced by 0.26 dex, with $\sigma=0.28$K~km~s$^{-1}$) with the fraction of clean emission (dimensionless, in color scale) corrected for the error beam contribution,  obtained after the two-step iterative process described in Sect~\ref{Error-beam}. The  beam size (11.2$\arcsec \times$ 11.2 $\arcsec$) is represented by the blue-filled circle at the bottom-left corner of the figure.}
\end{figure*}


\begin{figure}[ht!]
\centering
\includegraphics[width=1\linewidth]{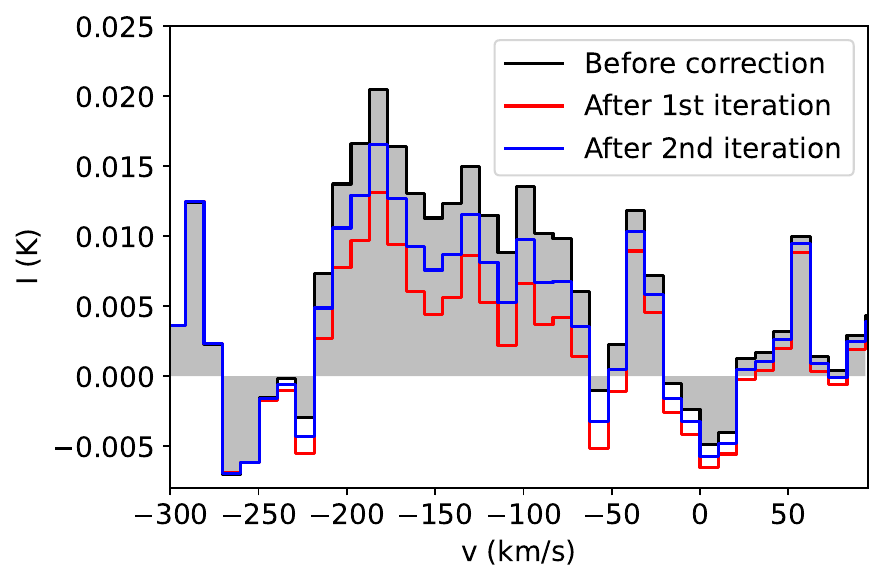}
\caption{\label{fig:spectrum-outplane} Spectrum extracted at  ($\Delta x$, $\Delta z$)~=~(--75$\arcsec$, --30$\arcsec$). The different curves show the initial spectrum before the correction (black) and the result obtained after the first (red curve) and second (blue curve) iteration of the error beam estimation algorithm described in Sect~\ref{Error-beam}.}
\end{figure}


\section{Error beam estimation}\label{Error-beam}

As we aim to characterize the vertical extent of the molecular gas in NGC\,891, it is of paramount importance to correct the observed signal for contributions from the IRAM 30m error beams. 
The telescope response to a point source is not represented by a single, ideal Gaussian profile. Instead, it includes additional broader, low-amplitude contributions which can add up to 30-40\% of the total power at 1-3mm \citep{Greve_2009, Kramer13}.

Let $T$ be the real brightness temperature distribution in the sky in a certain velocity channel. The \lancom{ideal} main beam brightness temperature ($T_{\mathrm{mb}}$) is defined as

\begin{center}
\begin{equation}
T_{\mathrm{mb}} = G_{\mathrm{mb}} \otimes T\lancom{,} 
\end{equation}
\end{center}

\noindent
where $\otimes$ indicates convolution over the sky and $G_{\mathrm{mb}}$ is the Gaussian function of unit area that represents the telescope main beam. In practice, however, the \lancom{measured} main brightness temperature delivered by the telescope ($\hat{T}_{\mathrm{mb}}$) has an extra contribution from the emission captured by the broader error beams. As a result,

\begin{eqnarray}
    \hat{T}_{\mathrm{mb}} = T_{\mathrm{mb}} + Q\otimes T,~\mathrm{with}\\    
    Q = \sum_{i=1}^3\dfrac{P'_i}{B_{\mathrm{eff}}} {G}_i.
\end{eqnarray}

\noindent
Here, the $G_i$ functions represent the three Gaussian error beams of the 30m telescope characterized by \citet{Kramer13},  $B_{\mathrm{eff}}$ is the main beam efficiency, and  $P'_i$ is the beam efficiency of the $i^\mathrm{th}$ beam error \red{(see Table~\ref{tab:errorbeam_params})}. For observations at 230~GHz with the 30m telescope, the $P'/B_{\mathrm{eff}}$ ratios are 7, 18 and 18$\%$, respectively, meaning that their contribution can be typically neglected. However, when observing faint emission (e.g., the halo) close to an extended and brighter region (e.g., the disk), this contamination must be assessed and removed.  

In this paper, we adopt the correction method devised by \citet{Garcia-Burillo92}, which we summarize below. In short, we define the following iterative approach: 

\begin{equation}
T_{\mathrm{mb}}|_N \equiv 
\begin{cases}
\, \hat{T}_{\mathrm{mb}}, & N=0\\[0.2cm]
 \, \hat{T}_{\mathrm{mb}}-Q\otimes T_{\mathrm{mb}}|_{N-1}, & \forall N>0.    
\end{cases}
\end{equation}

\noindent
Here, $T_{\mathrm{mb}}|_N$ is the $N^\mathrm{th}$ corrected main brightness temperature, with the $0^\mathrm{th}$ value corresponding to the actual measurement. It can be shown that, for $N>0$,  

\begin{equation}
   T_{\mathrm{mb}}|_N = \hat{T}_{\mathrm{mb}} + \left(\sum_{\alpha = 1}^N(-Q)^\alpha\right)\otimes\hat{T}_{\mathrm{mb}}.
\end{equation}

Using that $\hat{T}_{\mathrm{mb}} = T_{\mathrm{mb}} + Q\otimes T$,

\begin{equation}
   T_{\mathrm{mb}}|_N = {T}_{\mathrm{mb}} + \left(\sum_{\alpha = 1}^N(-Q)^\alpha\right)\otimes{T}_{\mathrm{mb}} - \left(\sum_{\alpha = 1}^{N+1}(-Q)^\alpha\right)\otimes T.
\end{equation}

Regrouping terms, we can rewrite this as 

\begin{equation}
   T_{\mathrm{mb}}|_N = T_{\mathrm{mb}} - (-Q)^{N+1}\otimes T_{\mathrm{mb}} +  \left(\sum_{\alpha = 1}^{N+1}(-Q)^\alpha\right)\otimes\left(T_{\mathrm{mb}}-T\right). 
\end{equation}
 
The equation above implies that the $N^{th}$ corrected value is equal to the actual main brightness temperature plus two correcting terms. The first one, \mbox{$- (-Q)^{N+1}\otimes T_{\mathrm{mb}}$}, accounts for the major contribution. This term changes its sign in each iteration and, being of order $(P'/B_{\mathrm{eff}})^{N+1}\ll1$, it becomes progressively smaller. The second term is \lancom{a} wavy distribution of order $P'/B_{\mathrm{eff}}$ with a null contribution to the area-integrated total flux. In practice, since all the error beams are much broader than the main one, $Q\otimes T_{\mathrm{mb}}\approx Q\otimes T$, so this term is negligible. 
For the purposes of this paper, we have applied this iterative method up to $N=2$, thus we consider $T_{\mathrm{mb}}|_2$ our best estimate of the main brightness temperature corrected for error beam effects.

As the Gaussian kernels must extend over an area that goes beyond the region mapped in the observations, we first create a symmetrized mock image of the galaxy. In this image, we assign the missing pixels on the western side of the galaxy $(x', z, v')$ a value of the CO flux $T_{\rm mb [x'-x_{\rm c}, z-z_{\rm c}, v'-v_{\rm sys}]}=T_{\rm mb [x_{\rm c}-x', z-z_{\rm c}, v_{\rm sys}-v']}$, where ($x_{\rm c}, z_{\rm c}$) and ${\rm v}_{\rm sys}$ correspond to the coordinates of the dynamical center and the systemic velocity of the galaxy, respectively \citep[${\rm v}_{\rm sys}^{{\rm HEL}}=535$~km~s$^{-1}$;][]{Garcia-Burillo92}.

Figure~\ref{fig:contamination} shows the observed CO velocity-integrated intensities ($I_{\rm CO}$) obtained by integrating the emission arising from the full velocity range due to rotation in the mapped region, ${\rm v}-{\rm v}_{\rm sys}=[-300, 300]$~km~s$^{-1}$, before any correction. This emission is overlaid, as white contours, on the color map of the estimated 
fraction of cleaned CO emission $\frac{I_{\rm corrected}}{I_{\rm uncorrected}}$, obtained after correcting for the error beam contribution using the two-step process described above.  The percentage of  emission free of contamination by the error  beams is high ($\geq 90\%-95\%$) across the disk up to vertical distances $\Delta z = \pm 18\arcsec$. Although this percentage decreases, it nevertheless remains in the range of $70-90\%$ 
at $-30\arcsec\leq\Delta z\leq25 \arcsec$ across most of the radial slices,
a result that confirms the detection of spatially resolved molecular gas emission up to vertical distances of about $1-1.5$~kpc. 
By analyzing the relative variation between the first and second iteration steps in each point of the integrated intensity map, we estimate that the correction converges to within an uncertainty of $\sim$7\%. These results agree with the error beam contribution to the CO emission in M\,51 obtained by \citet{Jakob22}.
For the rest of this work, we use the corrected version of the CO map.

Figure~\ref{fig:spectrum-outplane} illustrates the oscillatory nature of the solution found by the two-step process adopted in this work.
The figure compares the three versions of the CO emission detected at a representative high vertical distance above the galaxy 
midplane ($\Delta x$, $\Delta z$)~=~(--75$\arcsec$, --30$\arcsec$), corresponding to the uncorrected emission, the emission obtained after the first iteration, and the final version of the spectrum after the second iteration. 


\begin{figure*}[htbp]
\centering
\includegraphics[width=0.9\linewidth]{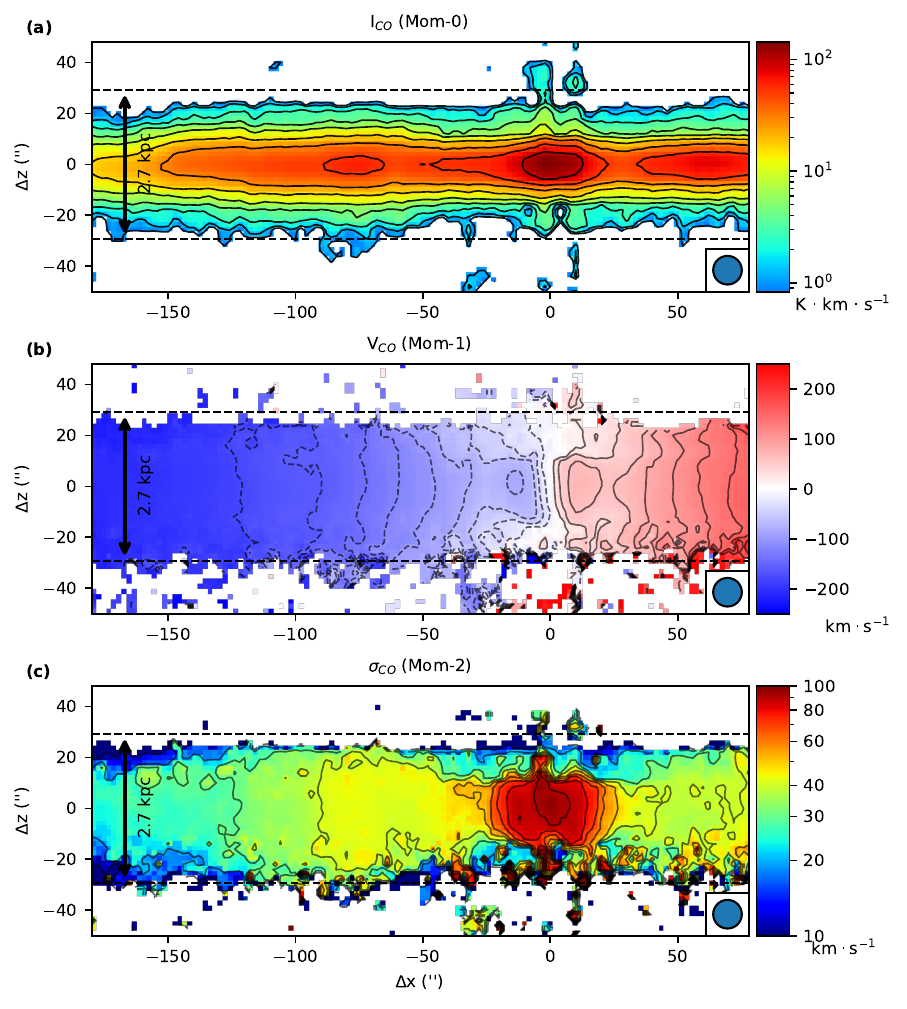}
\caption{\label{fig:moments_co} CO moment maps. From top to bottom: \lancom{velocity-integrated} intensities ({\em panel \lancom{a}}),  \lancom{intensity-weighted} velocities ({\em panel b}) and \lancom{velocity} dispersions ({\em panel c}) derived from the CO(2--1) line in the mapped region. Contour levels for the CO intensities increase logarithmically from 3$\sigma$ to 600$\sigma$  in steps of 0.26 dex, with $\sigma=0.28~ $K~km~s$^{-1}$. Mean-velocity contours range from -180 to 180~km~s$^{-1}$ in steps of 20~km~s$^{-1}$. Dashed lines indicate negative contour values. Dispersion contours go  from 10 to 100~km~s$^{-1}$ in steps of 10~km~s$^{-1}$. The 30~m beam size (11.2\arcsec) is indicated as a filled circle in the bottom left corner of each panel. A vertical thickness of 2.7\,kpc is represented as horizontal dashed lines, as visual reference.}
\end{figure*}


\begin{figure}[ht!]
\centering
\includegraphics[width=1\linewidth]{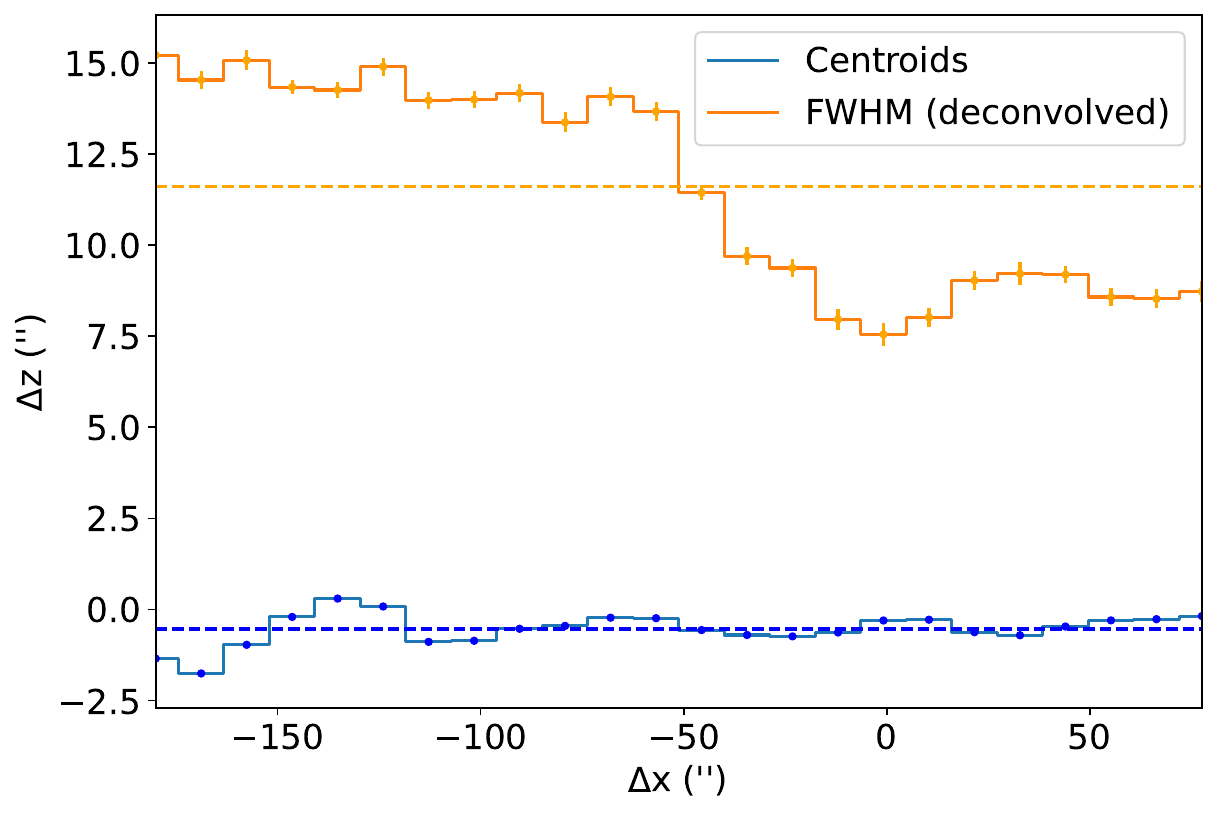}
\caption{\label{fig:fwhm_centroids} 
\lancom{Centroids} (blue histogram) and deconvolved FWHM thickness values (orange histogram)  derived from the single-Gaussian fits of  the out-of-plane molecular gas distribution across the major axis of the galaxy. \lancom{Dashed} lines show the estimated mean values of $\Delta z$ and FWHM.
The error bars are derived from the covariance of the parameters from the Gaussian fit.
}
\end{figure}

\begin{figure*}[htbp]
\centering
\includegraphics[width=0.85\linewidth]{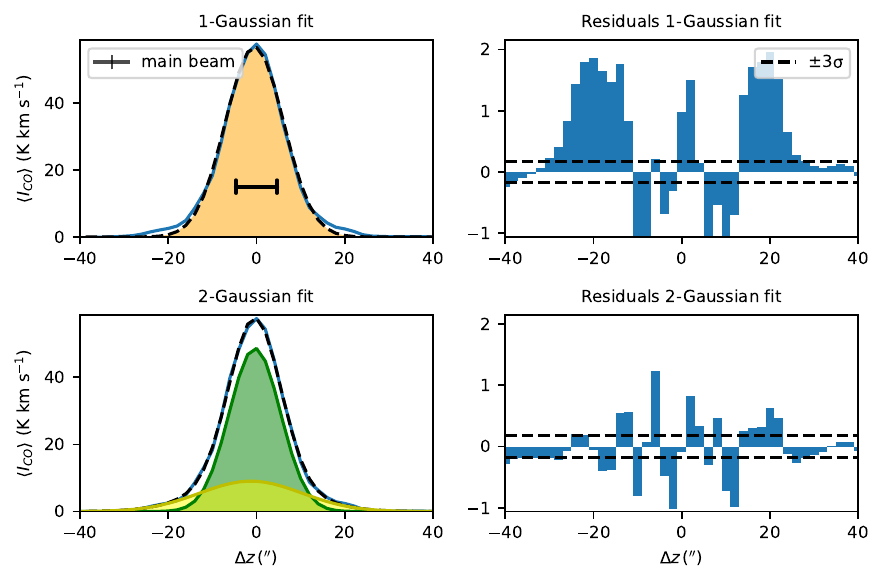}
\caption{\label{fig:gauss_res_co}
Comparison of the average out-of-plane distribution of CO\lancom{, <$I_{\rm CO}$>($\Delta z$),} with the \lancom{single-Gaussian fit} ({\em upper panels}) and \lancom{double-Gaussian fit} ({\em lower panels}) decomposition models described in Sect.~\ref{CO-decomposition}.  The blue curve shows the observed profile <$I_{\rm CO}$>($\Delta z$). The green and yellow curves in the {\em lower left panel} show the fit of the thin and thick molecular disk components of the double-Gaussian fit, respectively. The black dashed curves represent the sum of all the components in both models.  The {\em upper} and {\em lower right panels} display the residuals of the fits. Horizontal dashed lines indicate the $\pm3\sigma$ levels with $\sigma$ = 58 mK~km~s$^{-1}$. \red{The size of the main beam is indicated at the top left corner of the first panel.}}
\end{figure*}


\section{Results}\label{res_disc}
\subsection{CO moment maps}\label{COmoment}

Figure~\ref{fig:moments_co} shows the zeroth, first\lancom{,} and second-order moment maps of the CO(2-1) line obtained in the disk and disk-halo interface region of NGC~891, which is mapped in this work. These moment maps were obtained from the data cube version corrected for the error beam contribution, as described in  Sect.~\ref{Error-beam}. To derive the zeroth moment map of CO\lancom{, $I_{\rm CO}$($\Delta x$ ,$\Delta z$)}, we integrated the emission over the velocity range $\Delta$v$_{\rm tot}$~=${\rm v}-{\rm v}_{\rm sys}=[-300, 300]$~km~s$^{-1}$,  encompassing the full extent of velocities due to rotation, as shown in the CO channel maps of Fig.~\ref{fig:chmap_co_ha} and \ref{fig:chmap_co_usm}. The first and second-order moment maps were derived by applying a 3$\sigma$ clipping to the emission within $\Delta$v$_{\rm tot}$.    

While most of the total CO emission ($\simeq$70$\%$) stems from a thin, edge-on disk within a range of vertical 
distances of $\Delta$z$\simeq$~$\pm$10$\arcsec$ ($\pm$475~pc), a visual inspection of Fig.~\ref{fig:moments_co} shows the detection of 
statistically significant ($\geq$3$\sigma$) CO emission extending to a larger range of vertical distances above and below the galaxy midplane at z=0$\arcsec$:   20$\arcsec$(1~kpc)~$\lesssim$~$\mid \Delta z \mid$~$\lesssim$~30$\arcsec$ (1.4~kpc). The lowest-level contours of Fig.~\ref{fig:moments_co}  reveal a filamentary pattern of the out-of-plane CO emission. 

The first moment map shows the \refcom{intensity-weighted} mean velocity field of molecular gas. As expected, the kinematics of the gas are dominated by the galaxy's rotation. Gas emission appears blueshifted (redshifted) for $\Delta$x$\leq$0$\arcsec$ ($\Delta$x$\geq$0$\arcsec$). The velocity gradient along the galaxy's major axis is steepest within $\mid \Delta x \mid$~$\leq$20$\arcsec$ due to the presence of high-velocity gas emission from the circumnuclear  disk (CND) associated with the $x_{\rm 2}$ orbits of the stellar bar \citep{sgb95}. A quantitatively similar signature of rotation is imprinted on the gas lying at high vertical distances. In particular, the isovelocity contours are mostly parallel to the galaxy's minor axis outside the CND. This pattern indicates that the out-of-plane molecular gas component shows no rotation lag, unlike the $\HI$ emission at large distances \citep{Oosterloo07}.  

The second moment map shows the measured velocity dispersion ($\sigma_{\rm v}$) of the CO line across the mapped region. 
The widest lines ($\sigma_{\rm v} >70$~km~s$^{-1}$) are found close to the center of the galaxy, where the steep velocity gradients 
associated with the CND combined with the high inclination of the disk cause the highest degree of beam smearing. Outside the CND, the velocity dispersion shows a monotonic decline  along the major axis. Nevertheless, the gradient of $\sigma_{\rm v}$  measured along the minor axis is significant only within the 
CND region. At larger radii the vertical gradient of $\sigma_{\rm v}$ is very shallow. CO lines have comparable widths at different 
vertical distances for a given $x$-offset. \lancom{Similar to} the CO spectra of the midplane, this suggests that the out-of-plane CO signal also stems from the sum of line emission arising at multiple velocity components along the line of sight.


\begin{figure*}[htbp]
\centering
\includegraphics[width=.85\textwidth]{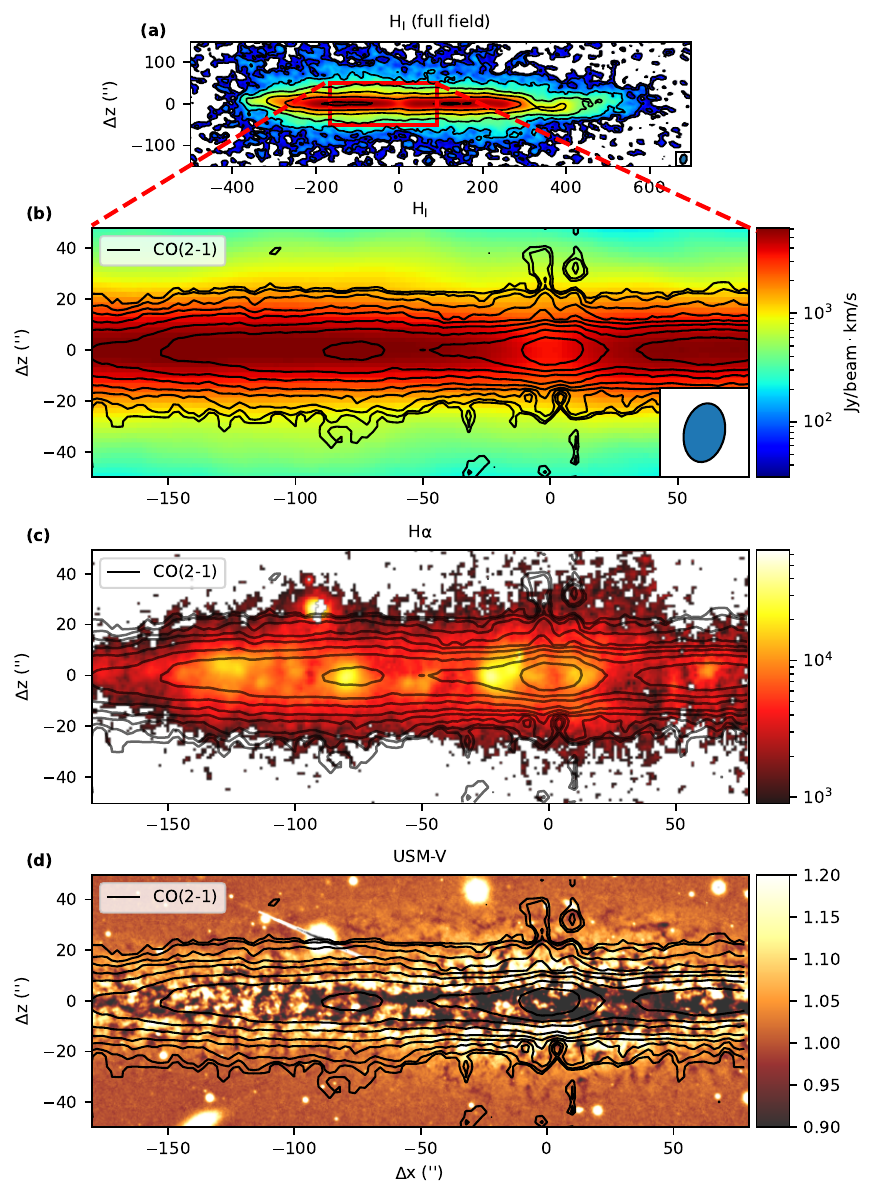}
\caption{\label{fig:tracers_ngc891_pel}
Comparison of the CO map with images obtained in other tracers of the ISM in NGC~891.
{\em Panel a} shows the $\HI$ map obtained by  \citet{Oosterloo07} from their full-resolution dataset (in logarithmic color scale and 
contours from 3\,$\sigma$ to 600\,$\sigma$ in steps of 0.38 dex, with $\sigma$ = 10.25~ Jy~beam$^{-1}$~km~s$^{-1}$). The blue-filled ellipse in the lower left corner of the panel represents the beam of the $\HI$ observations ($23.4\arcsec \times 16\arcsec@PA=12.3^{\circ}$).  The remaining panels, from top to bottom, show the overlay of the  CO intensities (in contours) with  (in color scale) the 
$\HI$ image ({\em panel b}), the  H$\alpha$ image from \citet{Kamp06}  ({\em panel c}, arbitrary units), and the  USM V-band image from  \citet{HS00}   ({\em panel d}) in the region of NGC~891 
mapped in this work. The CO contours in all the panels are  the same as in Figure~\ref{fig:moments_co}, covering the same field of view.}
\end{figure*}


\subsection{The thin and the thick molecular gas disks}\label{CO-decomposition}

The distribution and kinematics of molecular gas derived from the CO moment maps presented in Sect.~\ref{COmoment} provide supporting 
evidence for the existence of a significant out-of-plane molecular gas component extending  to vertical distances of $\Delta z\simeq1-1.5$~kpc.  In this section, we model the average out-of-plane distribution of CO emission using  Gaussian components to quantify both the vertical extent and mass of the thick molecular disk. Gaussian models are expected to provide a reliable description of the vertical structure of molecular gas under the assumption of hydrostatic equilibrium for the gas disk, which is embedded in the
larger-scale external potential of the stellar disk \citep[e.g.,][]{Becq1997}.

Figure~\ref{fig:fwhm_centroids} presents the results  of fitting the vertical CO distribution shown in Section~\ref{COmoment} as a function of the offset along the major axis with a single-Gaussian component. Fig.~\ref {fig:fwhm_centroids} shows the fitted $\Delta z$ centroids and the deconvolved 
FWHM values of the Gaussians as a function of the offset $\Delta x$ along the major axis. The Gaussian centroids cluster around a value of $\Delta z \simeq -0.5$\,\arcsec, which represents a small offset relative to the assumed major axis locus and to our spatial resolution.  This offset 
could be attributed to residual pointing errors of the 30m telescope. The centroids show no significant trend across the full radial extent of the mapped region. 
However, the evolution of the deconvolved FWHM values along the major axis shows a significant 
increase outside the central region: FWHM$_{[\mid \Delta x \mid >50 \arcsec]}\simeq 13.7\arcsec$ (640~pc)~$\simeq 1.6\times$ FWHM$_{[\mid \Delta x \mid <50 \arcsec]}\simeq 8.7\arcsec$(400~pc). 
 
Although the CO interferometer map of the galaxy published by \citet{Scoville93} shows no counterpart of the thick molecular disk, the reported thickness (FWHM) of the thin disk emission is also found to increase from the galactic nucleus ($\sim160$~pc) to the edge of the galaxy ($\sim 276$~pc).

\red{While the phenomenon of flaring disks is typically identified in $\HI$ and interpreted as resulting from the lower gravitational pull exerted on the gas by the shallower potential of the outer disk regions, the reported trend in the average thickness of the CO disk could be formally described in terms of a similar scenario for molecular gas. As shown in Figure~\ref{fig:fwhm_centroids}, the lowest\lancom{-}FWHM values are found at $\mid \Delta x \mid <50 \arcsec$, where the deeper gravitational potential of the galaxy center compresses the disk, thereby constraining the vertical extent of the molecular gas. Beyond the CND region ($\mid \Delta x \mid <50 \arcsec$) we do not find any statistically significant evidence of an increase in the FWHM values that could be  attributed to a flared disk at larger radii. \citet{Oosterloo07} also ruled out the idea of a flaring disk in $\HI$ with their models.} 

Figure~\ref{fig:gauss_res_co} shows the mean out-of-plane CO distribution in the mapped area of NGC~891\lancom{, <$I_{\rm CO}$>($\Delta z$)}, obtained by averaging $I_{\rm CO}$($\Delta x$ ,$\Delta z$)  along the major axis.  Figure~\ref{fig:gauss_res_co}  also 
presents the results of fitting the observed profile with  one and two Gaussian components. 
The parameters derived from the fits are listed in Table~\ref{tab:fits_parameters}. \refcom{The listed uncertainties of the different parameters correspond to formal fitting errors.}

The <$I_{\rm CO}$>($z$) profile shows evidence  of a thick molecular gas disk that extends  to vertical distances of $\mid \Delta z \mid \simeq$30$\arcsec$(1.4~kpc). The single-Gaussian fit of the  <$I_{\rm CO}$>($z$) profile of FWHM = 490~pc leaves large  residuals that are up to  25$\sigma$  ($\sigma$ = 58 mK~km~s$^{-1}$) over a wide range of vertical offsets 
from $\pm12\arcsec$ ($\pm$0.6~kpc) to $\pm30\arcsec$ ($\pm1.4$~kpc). The thickness of the disk determined by the single-Gaussian model is likely overestimated due to the presence of out-of-plane gas, which is inadequately described by the model.


\begin{figure*}[htbp]
\centering
\includegraphics[width=.95\linewidth]{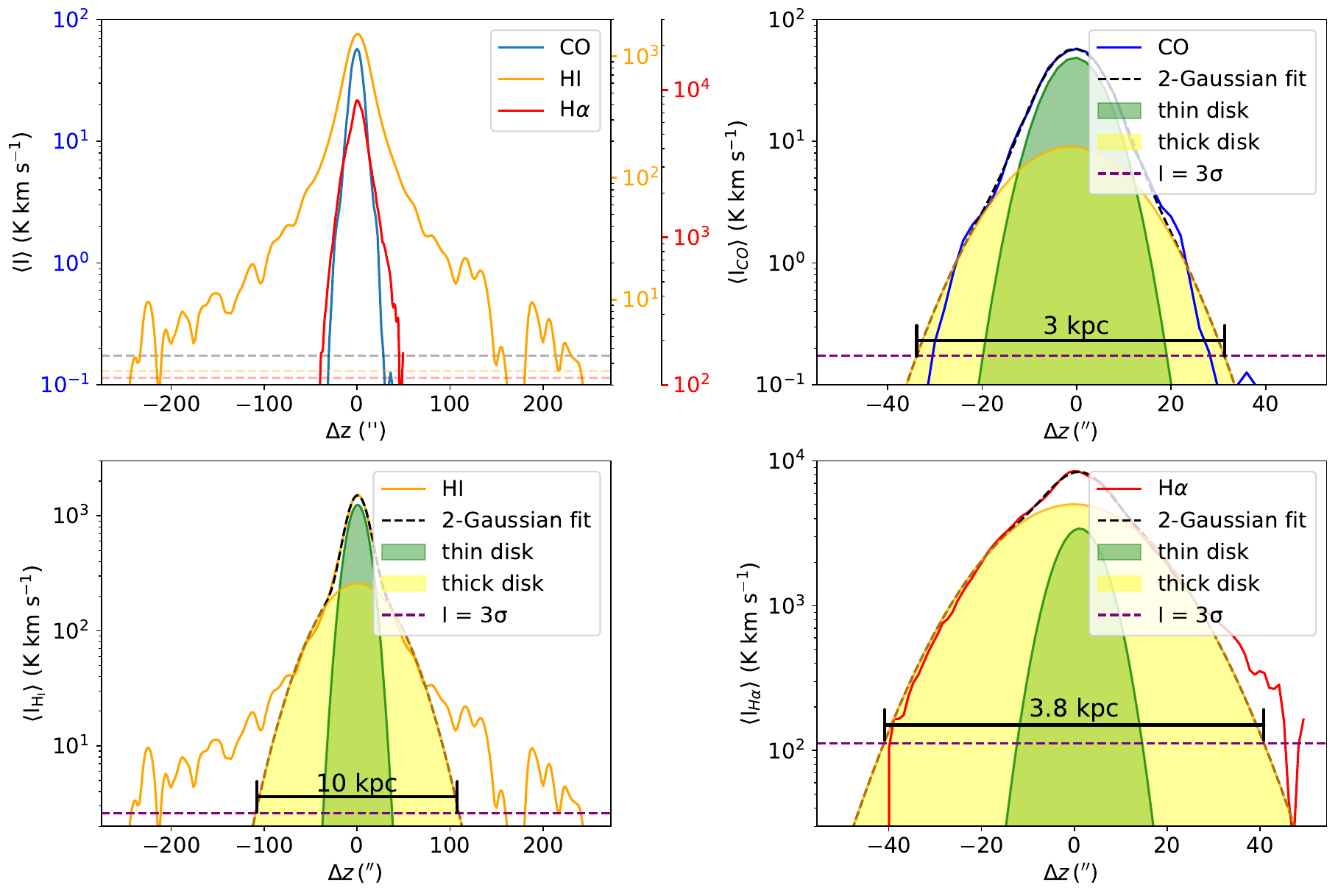}
\caption{\label{fig:perp_dist_co_ha_hi} 
\textit{Upper left panel}: Comparison of the average out-of-plane profiles of CO, $\HI$, and  H$\alpha$. All panels display the $y$-axis in logarithmic units. The integration ranges over the $x$-axis used to derive the profiles $[\Delta x_{\rm min}, \Delta x_{\rm max}]$ are: [-180$\arcsec$, 80$\arcsec$] for CO, [-1150$\arcsec$, 1050$\arcsec$] for $\HI$, and  [-180$\arcsec$, 80$\arcsec$] for H$\alpha$. \textit{Upper right panel}: Results of the double-Gaussian fit (black dashed line) of the CO profile (blue solid line). The thin disk and thick disk Gaussian components are shown in green and yellow lines, respectively. \refcom{Due to the current spatial resolution, the thin component fit should be taken as an upper limit}. \textit{Lower panels}: Same as  \textit{Upper right panel} but for the $\HI$ and H$\alpha$ profiles.  Horizontal dashed lines show the $3\sigma$ thresholds for each tracer in all panels ($\sigma_{\rm CO}=58$~mK~km~s$^{-1}$, $\sigma_{\rm \HI}=864$~mK~km~s$^{-1}$, and $\sigma_{\rm H\alpha}=37$~K~km~s$^{-1}$).}
\end{figure*}


The use of two Gaussian components provides a substantially improved representation of the out-of-plane CO distribution, consisting of a thin 
disk (FWHM = 360~pc) and a thick disk (FWHM = 1.1~kpc). 
\refcom{Since the thick disk is significantly broader than the main beam (516\,pc $\times$ 516\,pc), it appears to be well resolved at our current spatial resolution. However, the estimated vertical extent of the thin disk should be regarded as an upper limit to the intrinsic value due to the potential residual contribution of the extended thick disk component to the derived FWHM of 360\,pc.}
As shown in Fig.~\ref{fig:gauss_res_co}, the 
large residuals of the single-Gaussian model disappear at large vertical offsets in the double-Gaussian fit. 
The full extent of the thick molecular disk measured at a 3$\sigma$ level in  <$I_{\rm CO}$> is $\simeq 2.7$~kpc. Nonetheless, the double-
Gaussian fit overestimates  the value  of  <$I_{\rm CO}$> by a factor of $\simeq 2$ in the range of vertical offsets of $\mid \Delta z \mid \simeq1.3-1.5$~kpc. 

We estimate the fraction of molecular gas mass contained in the thick disk component relative to the total gas mass  ($f_{\rm thick}$)   
from the Gaussian decomposition described above. A conservative lower limit on $f_{\rm thick}\simeq3\%$ is obtained from 
the normalized flux of the residuals of the single-Gaussian model, although this likely underestimates  the true value. The double-Gaussian model estimates the value of $f_{\rm thick}$ to be up to $\simeq 27\%$. 
\red{Under the assumption that the same CO-to-H$_2$ ($\alpha_{\rm CO}$) conversion factor applies across the observed region,  we estimate that the thick disk can contribute up to $\rm\sim27\%$ of the total molecular mass. This fraction is similar to the estimated value of $\rm\sim35\%$ for the thick molecular disk imaged in the CO(1–0) line, as reported in \citet{Sof93}.
However, the assumption of a common $\alpha_{\rm CO}$ for both the thin and the thick molecular disks is debatable, given the dependencies of the conversion factor on physical parameters such as far-ultraviolet radiation \citep{Bolatto13, Heyer25}, gas density and kinetic temperature \citep{weib01}, and metallicity and ISM structure \citep{Bolatto13, Sch-Leroy24}. 
A detailed evaluation of the conversion factor in the thick molecular disk would require the observation of additional transitions of CO and its \lancom{isotopologs}.
Nevertheless, for the purposes of this work, we adopt a constant $\alpha_{\rm CO}$ to obtain first-order estimates of the relative molecular gas content in the thin and thick disk components\lancom{,} and to compare them with the values derived in other studies.}

\begin{table}[tbh!]
 \centering
 \small
 \caption{Results obtained from the single and double Gaussian fitting.}
\resizebox{\columnwidth}{!}{
\begin{tabular}{|l||c|c|c|}
 \hline
  & CO (2-1) & $\HI$ & H${\alpha}$ \\
 \hline
 A$_{\rm 1G}$ & 0.970 $\pm$ 0.008  & 0.823 $\pm$ 0.009 & 0.95 $\pm$ 0.01 \\
 FWHM$_{\rm 1G}$ [kpc] & 0.49  $\pm$ 0.01 & 1.49 $\pm$ 0.02 & 1.25 $\pm$ 0.02\\
 $\Delta z_{\rm 3\sigma}$ [kpc] & 2.07 & 4.45 & 3.12\\
\hline
 A$_{\rm 2G}$ & 1.008  $\pm$ 0.006 & 0.956 $\pm$ 0.004 & 0.995 $\pm$ 0.006 \\
 A$_{\rm thin}$ &  0.73 $\pm$ 0.04  & 0.586 $\pm$ 0.006 & 0.19 $\pm$ 0.01\\
 A$_{\rm thick}$ &  0.27 $\pm$ 0.04 & 0.414 $\pm$ 0.006 & 0.81 $\pm$ 0.01 \\
 FWHM$_{\rm thin}$ [kpc] & 0.36 $\pm$ 0.02 & 1.15 $\pm$ 0.01 & 0.53 $\pm$ 0.02\\
 FWHM$_{\rm thick}$ [kpc] & 1.1 $\pm$ 0.1 & 3.91 $\pm$ 0.07 & 1.60 $\pm$ 0.02\\
 $\Delta z_{\rm thick, 3\sigma}$ [kpc]  & 3 & 10 & 3.8\\
\hline
\end{tabular}
}

    \tablefoot{Upper half of the table correspond to the results of the single-Gaussian fitting (1G), while bottom half shows the results of the double-Gaussian fitting (2G) and their individual components (thin and thick distributions).
    A$_{\rm 1G, 2G}$ represent the fraction of the observed flux in the profiles that is accounted for by the fitted components, FWHM$_{\rm 1G, 2G}$ are the deconvolved full widths at half maximum of each component, $\Delta$z$_{\rm 3\sigma}$ are the full widths of the fitted components measured at a 3$\sigma$ level.}
    \label{tab:fits_parameters}
\end{table}

\subsection{Comparison with other ISM tracers}\label{ISM-comparison}

Sections~\ref{COmoment} and \ref{CO-decomposition} present strong evidence for the presence of a thick molecular disk in NGC~891. In this section\lancom{,} we compare the distribution of molecular gas derived from the 30m CO map  with that derived from other tracers of the ISM, namely the neutral atomic hydrogen ($\HI$), the diffuse ionized gas (DIG; H$\alpha$), and the dust, in order to identify possible counterparts of the thick molecular disk of NGC~891. 

Figure~\ref{fig:tracers_ngc891_pel} shows the velocity-integrated CO intensity contours overlaid on the $\HI$ zeroth-moment map from \citet{Oosterloo07}, 
\red{on the H$\alpha$ map from \citet{Kamp06}\lancom{,} and on the unsharp-masked (USM) V-band image from \citet{HS00},} over the region of the galaxy mapped in this work.

The $\HI$ emission, derived from the high-resolution 
\red{observations from \citet{Oosterloo07} 
(data set available on the HALOGAS survey, \citet{halogas_2011})}, extends to large vertical distances far beyond the molecular component. At vertical distances of $\Delta z \gtrsim 6$~kpc, namely a factor of 4 larger than the maximum extent of CO in the thick disk, $\HI$ remains detectable. The low-resolution $\HI$ data of the galaxy of  \citet{Oosterloo07} reveal filamentary structures of the gas that reach heights of up to $\sim 22$~kpc. For the purposes of comparison with the CO data, we focus on the high-resolution $\HI$ map, whose beam size (FWHM$_{\rm \HI, HR} = 23.4'' \times 16''$) is more comparable to that of the CO observations (FWHM$_{\rm CO} = 11.2'' \times 11.2''$). At $\Delta z \sim 1.5$~kpc, where CO emission drops below the detection threshold of the observations, the $\HI$ emission remains fairly uniform across galactocentric radius. 

The DIG distribution, traced by H$\alpha$ emission from \citet{Kamp06}, shows a total vertical extent comparable to that of the molecular gas. The same conclusion can be drawn 
by comparing the CO map with the  H$\alpha$ image published by \citet{HS00}.
The H$\alpha$ emission exhibits an asymmetric distribution within the plane of the galaxy along its major axis, with comparatively stronger emission observed on the northeastern 
(approaching) side of the disk. This asymmetry is attributable to the location of the $\ion{H}{ii}$ regions, which are positioned  inside corotation, downstream from the spiral arm on the 
approaching side of the galaxy's disk \citep{Garcia-Burillo92}. Conversely, $\ion{H}{ii}$ regions \lancom{in} the southwestern side of the disk would suffer from the high extinction associated with the 
spiral arm. \lancom{We detect significant extraplanar emission} above and below regions of enhanced \lancom{SF}. Both CO and H$\alpha$ reach similarly large vertical distances near 
the galactic center and on the northeastern side of the disk, suggesting a close connection between local \lancom{SF} activity in the disk and the vertical extent of the gas.

To quantify the vertical structure of the different ISM tracers in NGC~891, we compared the performance of single-Gaussian (thin disk only) and double-Gaussian (thin+thick disk) models in reproducing the averaged vertical profiles of $\HI$, and H$\alpha$ emission, adopting a methodology similar to that used to analyze the CO data in Sect.~\ref{CO-decomposition}. The resulting best-fit parameters are summarized in Table~\ref{tab:fits_parameters}, and the corresponding fits are shown in Fig.~\ref{fig:perp_dist_co_ha_hi}. The molecular and ionized gas exhibit remarkably similar vertical extents, while the $\HI$ distribution extends significantly to larger vertical distances farther into the halo.

The improvement achieved by adopting a double-Gaussian model to reproduce the vertical structure of $\HI$ and H$\alpha$ emission is most clearly seen in the residuals (see Fig.~\ref{fig:gauss_res_ha_hi}). The CO and DIG maps exhibit a similar vertical structure, with detectable emission extending to $\Delta z_{3\sigma}^{\rm H\alpha} \simeq 3.8$~kpc. The $\HI$ gas shows the most extended distribution, characterized by a thin disk with FWHM $\sim 1.1$~kpc and a thick disk with FWHM $\sim 3.9$~kpc, reaching $\Delta z_{3\sigma}^{\rm \HI} \simeq 10$~kpc. 
\refcom{The beam size of the $\HI$ observations (1.73\,kpc $\times$ 1.07\,kpc) implies that the thin disk thickness should also be interpreted as an upper limit. In addition, alternative fitting models (e.g.\lancom{,} double-exponential models) may help to improve the description of the more diffuse thick disk component.}
\cite{Yim_2011} also concluded that the inclusion of a  second thick Gaussian component improves the fit of the observed vertical structure of the $\HI$ gas. These results highlight the progressively increasing vertical extent of the ISM phases from molecular and DIG to atomic gas.

\subsection{Kinematics: \lancom{Position}-velocity diagrams} \label{PV-diagrams}

\begin{figure}[ht]
\centering
\includegraphics[width=1\linewidth]{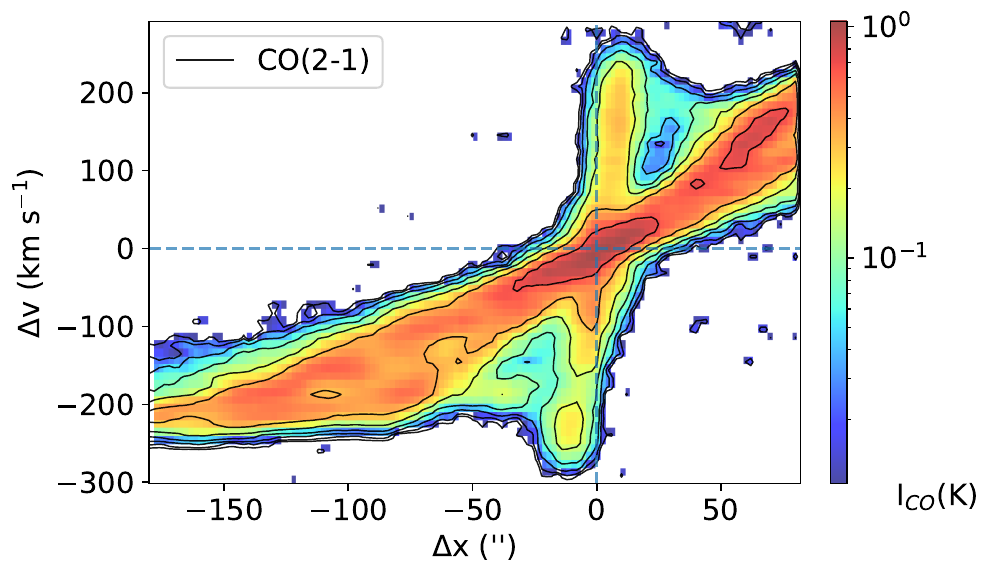}
\caption{\label{fig:pv_co} \lancom{Position-velocity} diagram of the CO(2-1) emission along the major axis at $\Delta$z=0. Black line contours are 3$\sigma_{\rm ch, CO}$, 6$\sigma_{\rm ch, CO}$, 12$\sigma_{\rm ch, CO}$, 24$\sigma_{\rm ch, CO}$, 48$\sigma_{\rm ch, CO}$, 96$\sigma_{\rm ch, CO}$ and 192$\sigma_{\rm ch, CO}$ ($\sigma_{\rm ch, CO}$ = 3.5\,mK). The blue dashed lines indicate the location of the dynamical center in the parameter space of the PV diagram.}
\end{figure}

To investigate the kinematics of the molecular gas, we constructed a position--velocity (PV) diagram of the CO(2--1) emission along the major axis of NGC~891 at $\Delta$z = 0 (Fig.~\ref{fig:pv_co}). The PV diagram clearly reveals the molecular ring and the fast-rotating inner nuclear disk. The latter is confined to the region $-15'' < \Delta x < +15''$ and reaches velocities of up to $ \mid {\rm v-v_{sys}} \mid \simeq 295$~km~s$^{-1}$. A gap separates this central component from the \lancom{large scale} disk rotation.  \citet{Garcia-Burillo92} and \citet{sgb95} interpreted the X-shaped morphology of the CO \lancom{PV} diagram as evidence that a stellar bar drives the gas flow. In the model of \citet{sgb95}, the bar has a corotation radius of $r\simeq70\arcsec$ (3.3~kpc) and is viewed at an angle $\alpha\simeq45^{\circ}$ from the galaxy's major axis. In this scenario, the molecular gas populates the $x_{\rm 2}$ orbits within a fast-rotating nuclear disk. This disk lies between the two Inner Lindblad Resonances \lancom{(ILR)} of the bar. Outside the nuclear disk, most of the CO emission in the PV diagram is associated with $x_{\rm 1}$ orbits within the bar up to $\Delta x = \pm 70\arcsec \times {\rm cos}(\alpha) \simeq \pm 50\arcsec$, as well as with the spiral arms that develop outside the bar at larger radii.


\begin{figure*}[htbp]
\centering
\includegraphics[width=0.85\linewidth]{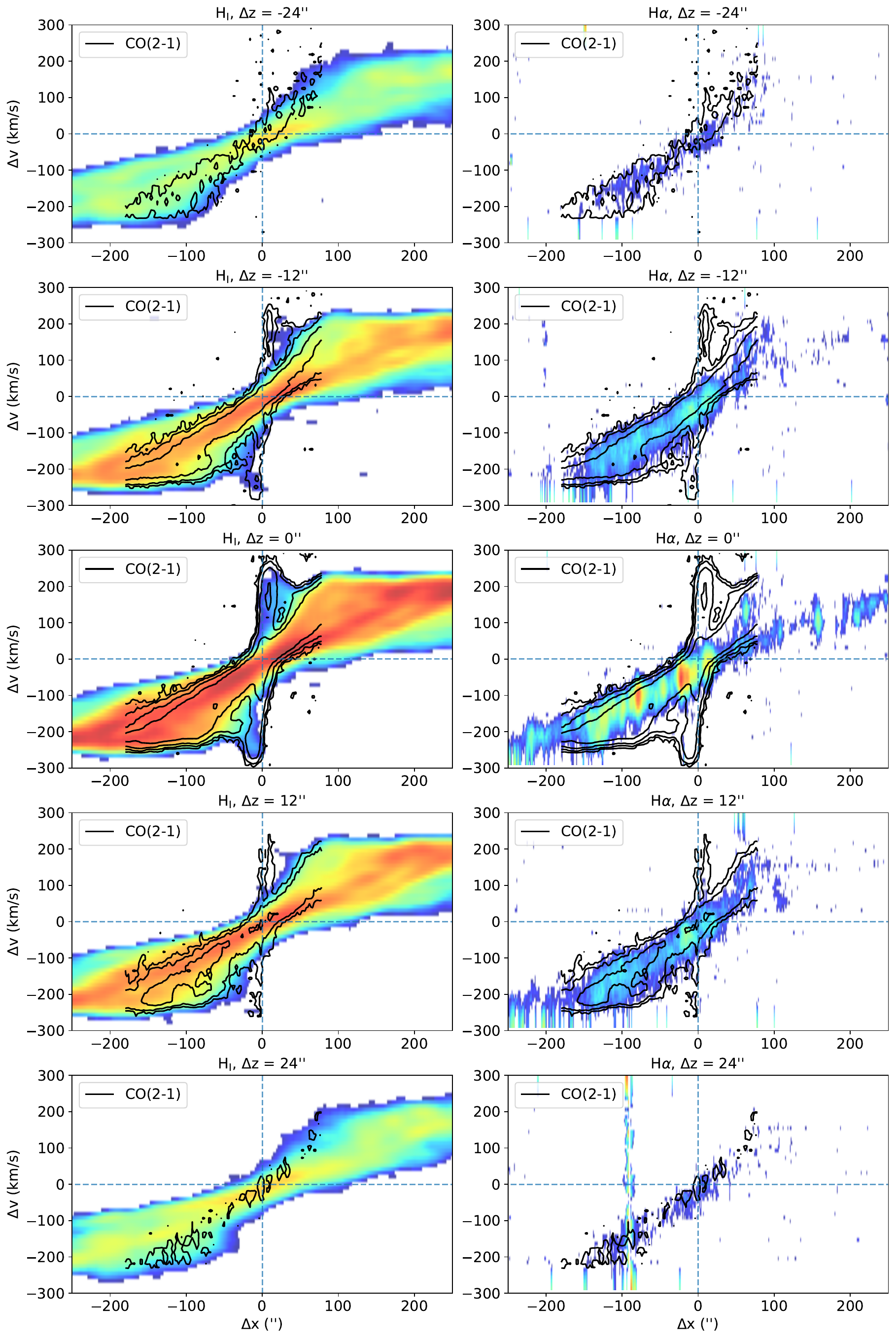}

\caption{\label{fig:pvx_hi_ha}\lancom{Position-velocity} slices parallel to the galactic plane of the $\HI$ (left) and H$\alpha$ (right) emissions. Slices are taken at $\Delta$z = 0\,\arcsec, 12\,\arcsec and 24\,\arcsec above and below the midplane. 
Color scales are clipped at 3\,$\sigma_{\rm ch}$ ($\sigma_{\rm ch, \HI}$ = 147\,mK, $\sigma_{\rm ch, H\alpha}$ = 3.8\,K). \lancom{Position-velocity} slices of the CO(2-1) emission are plotted as black contours, with levels 3$\sigma_{\rm ch, CO}$, 9$\sigma_{\rm ch, CO}$, 27$\sigma_{\rm ch, CO}$ and 81$\sigma_{\rm ch, CO}$ ($\sigma_{\rm ch, CO}$ = 3.5\,mK). Dashed blue lines denote the zero-offset coordinates.}
\end{figure*}

Figure~\ref{fig:pvx_hi_ha} compares the CO PV diagram with those derived from $\HI$ and H$\alpha$ emission. We extracted PV slices parallel to the midplane at different $\Delta z$ offsets. 
The H$\alpha$ PV diagram is dominated by a solid-body rotation pattern, consistent with previous results by \citet{Kamp06}.
Notably, there is no H$\alpha$ emission at high velocities in the nuclear disk. 
In addition, the H$\alpha$ PV diagram shows a marked east-west asymmetry. Both characteristics can be attributed to the strong extinction in the galaxy's midplane. 
In contrast, the CO emission shows prominent high-velocity features in the nuclear disk, which progressively disappear at larger vertical distances. Outside the inner region, the overall shape of the CO PV diagram transitions \lancom{toward} a structure similar to that of the DIG at higher $\Delta z$.
\red{Nevertheless, since CO is not affected by extinction, a direct comparison between the two tracers is not straightforward.}
The $\HI$ emission extends to larger radial distances, particularly on the southwestern side, but exhibits a less pronounced high-velocity component in the nuclear disk. This fast-rotating component rapidly vanishes in slices above and, more noticeably, below the midplane, indicating that it is confined to the thin inner disk, as noted by \citet{Oosterloo07}.

\begin{figure*}[htbp]
\centering
\includegraphics[width=0.95\linewidth]{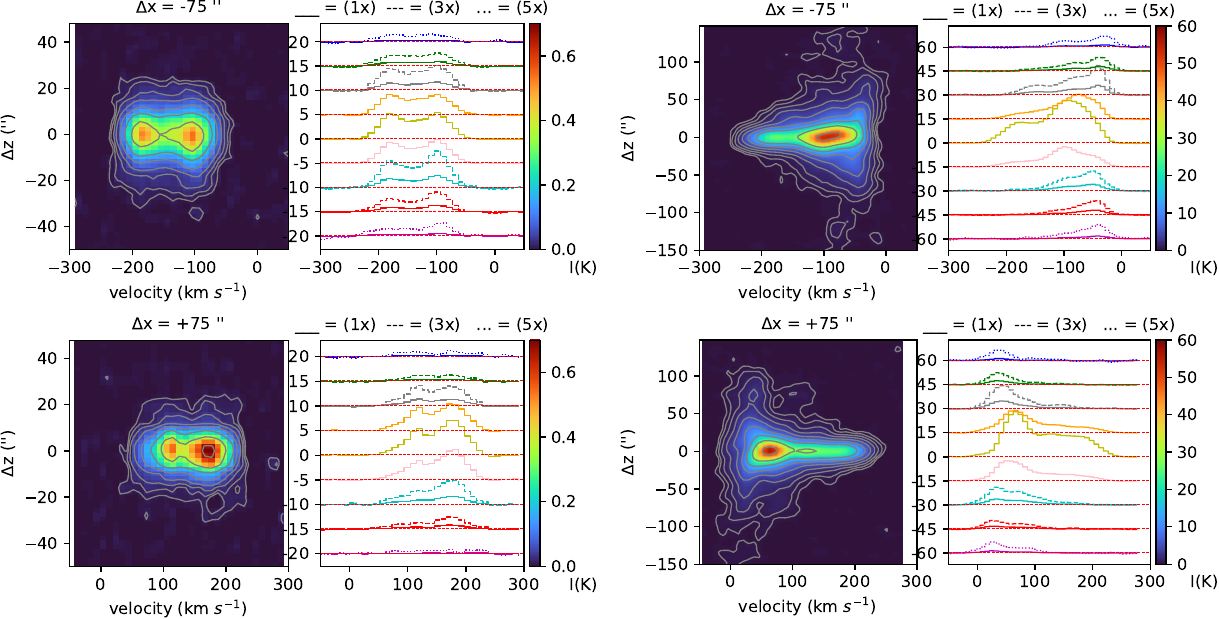}
\caption{\label{fig:pv_spec_sample} Position-velocity vertical slices of CO ({\em left panels}) and $\HI$ ({\em right panels}) at $\Delta x$ = $\pm$75$\arcsec$. Contours increase logarithmically from $3\sigma_{\rm ch}$ to $600\sigma_{\rm ch}$ ($\sigma_{\rm ch, CO}$ = 3.5 mK, $\sigma_{\rm ch, \HI}$ = 148 mK) in steps of 0.26 dex. Next to the PV slices we show the CO and $\HI$ spectra observed at different vertical $\Delta z$ offsets. Some of the spectra have been rescaled for clarity. The red dotted horizontal line denotes the $3\sigma_{\rm ch}$ detection limit.}
\end{figure*}

To explore the vertical dependence of the kinematics, we also extracted vertical PV slices at fixed radial offsets and examined spectra at different heights above the midplane. Figure~\ref{fig:pv_spec_sample} shows representative slices at $\Delta x = \pm 75''$ for CO and $\HI$, regions well beyond the galactic center. 
\red{The $\HI$ spectra show a systematic shift of the velocity peaks toward lower absolute velocities at high $|\Delta z|$, in agreement with the results obtained in the lagging models of \citet[][]{Oosterloo07} ($\rm \sim$15\,km\,s$^{-1}$\,kpc$^{-1}$). A similar velocity gradient was found for H$\alpha$ in \citet{Kamphuis07} ($\rm \sim$18.8 $\pm$ 6.3\,km\,s$^{-1}$\,kpc$^{-1}$). Nevertheless, our examination of the CO data failed to reveal any statistically significant indications of rotation lag beyond the central CND region, defined as $\mid \Delta x \mid <50 \arcsec$. The mean isovelocity contours are predominantly parallel to the galaxy's minor axis in the external regions, extending beyond the central nucleus. Furthermore, the \lancom{PV} cuts of Figures \ref{fig:pvx_hi_ha} and \ref{fig:pv_spec_sample} confirm the absence of any significant rotation lag signature in CO. The potential bias due to extinction, affecting H$\alpha$ but not CO, could explain the observed discrepancies in the kinematics of CO and H$\alpha$.}
At $\Delta x = \pm 70 \arcsec$, additional intensity peaks appear \red{on the CO maps} at lower velocities than the terminal velocity at these radii. These features likely correspond to prominent spiral arm segments that develop outside the bar. The kinematic signature of the spiral arm becomes visible near $\Delta x \simeq \pm 70''$,  subsequently merging with the terminal velocity of the main disk at larger radii. Notably, the vertical extent of CO emission shows little dependence on velocity. This suggests that the vertical extent of molecular gas is comparable inside and outside the spiral arms.
 
Figs.\ref{fig:pv_spec_CO}, \ref{fig:pv_spec_HI}, \& \ref{fig:pv_spec_Ha} show additional CO, $\HI$, and $H\alpha$ PV slices in Appendix~\ref{multiple-slices}\lancom{.}


\section{Discussion}\label{Discussion}

\subsection{The thick disk: Effect of inclination on the observed vertical extent}

Although the literature reports values for the inclination of the disk of NGC~891 \citep[$i \simeq 88-89^{\circ}$][]{Rup91, Garcia-Burillo92, Sof93} indicating an edge-on orientation of the galaxy, it is important to assess whether the projection effects of a tilted disk could artificially produce or enhance the appearance of a vertically extended component of molecular gas. In particular, the emission from gas at different galactocentric radii along the line of sight could mimic an increased vertical thickness.

In order to evaluate this effect, we constructed a set of disk models with different inclinations using the \lancom{3D-Based Analysis of Rotating Object via Line Observations (3D-BAROLO)}  package \citep{barolo15}. To this end, we assumed a simple geometrical model of a purely thin molecular disk with a thickness comparable to that of the Milky Way (FWHM $\sim$ 250~pc). As rotation curve\lancom{,} we adopted the one used by  \citet{{Garcia-Burillo92}} to fit the kinematics of molecular gas along the galaxy's major axis:

\begin{equation}
v_{\rm rot} = v_{\rm max} \cdot \sqrt{\frac{x^2}{x^2+d^2}} \lancom{,}
\label{eq:Tmb_step_2}
\end{equation}

\noindent where d = 200\,pc and ${\rm v}_{\rm max}$ = 225\,km$\cdot$s$^{-1}$ is the maximum rotational velocity. 

\begin{figure*}[htbp]
\centering
\includegraphics[width=0.88\linewidth]{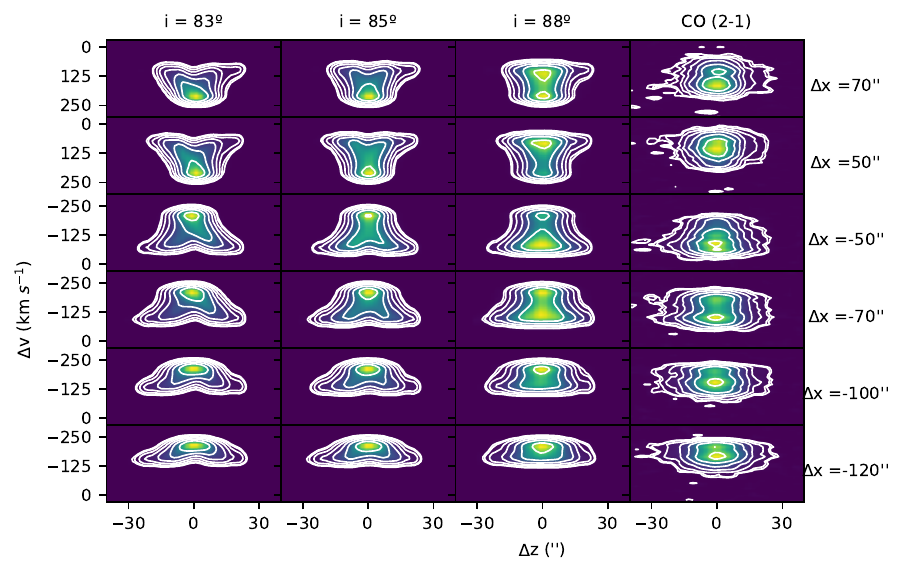}
\caption{\label{fig:bbarolo} \lancom{Position-velocity} vertical slices at different radial offsets $\Delta x$ and for different galaxy inclinations obtained by
 {\tt $^{3D}$BAROLO}. For each PV slice the horizontal axis is the $\Delta z$ offset in $\arcsec$ and the vertical axis is the velocity \lancom{toward} the observer in km~s$^{-1}$ relative to ${\rm v_{\rm sys}}$. The right-most column represents the PV slice that was obtained from the observations. White contours show the levels 3$\sigma_{\rm ch, CO}$, 4$\sigma_{\rm ch, CO}$, 8$\sigma_{\rm ch, CO}$, 16$\sigma_{\rm ch, CO}$, 32$\sigma_{\rm ch, CO}$, 64$\sigma_{\rm ch, CO}$, and 128$\sigma_{\rm ch, CO}$.}
\end{figure*}

Given the radial extent of the CO disk,  $\simeq 250\arcsec$ \citep{Garcia-Burillo92}, an inclination of $i\simeq83^{\circ}$ would be necessary to account for the observed vertical extent ($\simeq 30\arcsec$) of the thin disk (of $\simeq 5\arcsec \simeq 250$~pc-size) seen in projection. 
Figure~\ref{fig:bbarolo} shows the resulting simulated vertical PV plots taken at a subset of $\Delta x$ offsets along the major axis for values of the inclination $i=83^{\circ}$, $85^{\circ}$,  and $88^{\circ}$. The PV plots derived for $i=83^{\circ}$, the largest inclination that could explain the observed thick disk as an inclined thin disk, fail to reproduce the observed pattern of the CO emission. The PV plots for $i=83^{\circ}$ show that the emission of the gas at the outer edges of a rotating thin disk, seen in projection at $\Delta z\sim15\arcsec-30\arcsec$,  should be closer to the systemic velocity than the gas at  the major axis, seen at $\Delta z\sim0\arcsec$.  This velocity gradient is not consistent with the observations.
Even under extreme assumptions, projection effects alone cannot account for the detected emission at $|\Delta z| \gtrsim 1$~kpc. This result confirms that the identified thick molecular disk is not an artifact of inclination, but represents a genuine extraplanar component.

\cite{Swaters_1997} reached a similar conclusion regarding the inclination of the $\HI$ gas disk. While an inclination of $\simeq 80^{\circ}$ would be required to explain the observed $\HI$ thickness of the galaxy, a tilted disk at $<86^{\circ}$ produced a V-shaped signature in the PV plots that was  more pronounced than in the observations.


\subsection{The thick molecular disk and the galactic fountain scenario}

An order-of-magnitude estimate of the timescales involved in lifting molecular gas into the halo can provide further insight into the physical mechanisms 
that explain the observed properties of the thick disk.

The time required for molecular gas to reach the characteristic maximum height of $z_{\rm max} \sim 1.5$~kpc  derived from the observations, $t_{\rm lift}$, can be estimated if we assume a value for the kick vertical velocity ($v_{\rm kick}$).  \citet{fraternali06} modeled the extra-planar $\HI$  gas of NGC~891 assuming that gas clouds act as non-interacting particles lifted from the plane of the disk due to the expansion of superbubbles inflated by \lancom{SNe}. The velocity required to reproduce the distribution of the $\HI$ halo in \lancom{the} \citet{fraternali06} model is $\simeq 75$~km~s$^{-1}$. This value is in close agreement with the predictions of hydrodynamical models of \citet{Mac88}. Therefore, if  we assume $v_{\rm kick}=75$~km~s$^{-1}$: 

\begin{equation}
t_{\rm lift} \simeq \frac{z_{\rm max}}{v_{\rm kick}} \simeq 20~{\rm Myr}.
\end{equation}

This timescale is comparable to the lifetimes of massive stars and to the duration of active star-forming episodes ($t_{\rm SF}\simeq 10$~Myr), which supports the idea that stellar feedback can plausibly sustain the observed thick molecular disk. If the molecular gas subsequently cools and falls back toward the disk, the full fountain cycle would occur on timescales of a few tens of Myr.

Star formation activity in the disk is expected to be assembled along the spiral arms of the galaxy. In the event that the two-spiral arm structure of NGC 891 is coupled to the galaxy's stellar bar, it can be posited that both density wave features would exhibit an identical pattern speed, given by $\Omega_{\rm p} = \frac{{\rm v}_{\rm rot}}{R_{\rm cor}}$, where $R_{\rm cor}\simeq70\arcsec$ (3.3~kpc) is the bar's corotation radius, determined by \citet{sgb95}. The rotation of the \lancom{two-spiral-arm} pattern would propagate \lancom{SF} in the disk with an associated timescale ($t_{\rm rot}$) given by the following relation\lancom{:}

\begin{equation}
        t_{\rm rot} = \frac{1}{2}\frac{2 \pi}{\Omega_{\rm p}} \simeq 45\,{\rm Myr}\lancom{.}
\end{equation}

This is the timescale over which the two–spiral-arm pattern of NGC\,891 would sweep through the disk, producing mixing and triggering \lancom{SF} as gas passes through the $m=2$ density wave. The mixing induced by the spiral pattern on timescales of a few tens of Myr can account for the comparable vertical extent of molecular gas inside and outside the spiral arms, as discussed in Sect.~\ref{PV-diagrams}. The observed extraplanar CO emission along a given line of sight would therefore trace the cumulative effect of multiple ejection events over the past few tens of Myr.
This picture is further supported by recent studies showing that \lancom{SF} in disk galaxies is not 
\red{solely} confined to spiral arms, with a \red{non negligible} fraction occurring in interarm regions and contributing to the spatial diffusion and mixing of stellar feedback across the disk, in addition to the redistribution driven by disk rotation \citep[e.g.,][]{Foyle10, Kreckel16, Querejeta21, Querejeta24}.
Although these estimates are necessarily approximate and rely on simplified assumptions regarding the vertical velocity and gravitational potential, their similarity within a factor of a few demonstrates that the presence of a thick molecular component extending up to $\sim 1.5$~kpc is fully consistent with a galactic fountain scenario operating under typical disk conditions in NGC~891.

A detailed inspection of the CO channel maps  shown in Figs.\ref{fig:chmap_co_ha} \&  \ref{fig:chmap_co_usm} shows that the largest vertical extents of molecular gas occur predominantly in the northeastern half of the galaxy.  In several channels (e.g., ${\rm v}_{\rm ch} = -125$ and $-175$~km~s$^{-1}$), CO emission reaches heights of $\sim 40$--$45''$ (1.8--2.0~kpc) below the midplane. These regions coincide spatially with prominent dust filaments identified in the \lancom{USM} optical images of \citet{HS00} and with enhanced \lancom{DIG} emission traced by H$\alpha$. \citet{jwst24} used NIRCam imaging at 1.5 and 2.7~$\mu$m to reveal a tight connection between dust filaments and star-forming regions in the midplane of NGC~891, extending into the lower halo. Together with the CO and H$\alpha$ data presented here, these results indicate that the thick molecular disk is primarily shaped by feedback-driven outflows associated with disk \lancom{SF}.

The close spatial and kinematic correspondence between molecular gas, ionized gas, and dust strongly suggests that these components are linked by a common physical mechanism. In particular, the observed morphology is consistent with a galactic fountain scenario, in which stellar feedback from star-forming regions drives gas vertically from the disk into the halo. In this framework, molecular gas is entrained together with dust and ionized material, reaching moderate heights before cooling and returning to the disk.
The dust distribution, highlighted using the USM V-band image, reveals numerous dark filaments (\lancom{"}chimneys\lancom{"}) emerging from the midplane and extending to large vertical distances into the halo. A detailed comparison between the chimneys and the molecular gas features of the thick CO disk is not possible due to the significantly different spatial resolutions of these observations. However, the network of dark filaments is distributed over spatial scales comparable to those of the thick CO disk.
Several of these filaments are spatially coincident with regions of enhanced CO and H$\alpha$ emission. In particular, prominent structures near $\Delta x \simeq 100''$ and $140''$ extend to heights of $\sim 2$~kpc, comparable to the vertical extent of the molecular and ionized gas. Similar features were previously reported by \citet{HS00} and more recently by  \citet{jwst24}. 
A galactic fountain mechanism \citep{fraternali06} is likely responsible for most of the gas at heights of z$\leq$5\,kpc, whereas the outermost regions may contain $\sim$10\% of the halo gas, probably originating from accretion processes \citep{Oosterloo07}.

Alternative explanations involving external accretion appear less likely for the molecular component. While \citet{Oosterloo07} reported signatures of gas accretion in NGC~891 in the form of counter-rotating $\HI$ clouds and a large filament extending up to $\sim 22$~kpc from the disk, these structures occur at much larger heights than those probed by the molecular gas. The limited vertical extent of the thick molecular disk detected here suggests that inflow processes are not dynamically dominant at these heights.

\section{Summary and conclusions}\label{Conclusions}

We present new CO(2--1) observations of the nearby edge-on spiral galaxy NGC~891 carried out with the IRAM 30m telescope.  These observations provide unprecedented sensitivity and spatial coverage of the molecular gas distribution in the disk and the disk-halo interface of the galaxy. This project builds on the previous work of \citet{Garcia-Burillo92} who found evidence for CO emission extending over  1\,kpc outside the galactic plane of NGC~891.

Our main conclusions can be summarized as follows:

\begin{enumerate}

\item
 We confirm the detection of a vertically extended molecular gas component or thick disk by robustly characterizing and correcting for the residual error-beam contamination in the observations using an iterative approach.
 
 \item  
The vertical distribution of molecular gas is best fit by a double-Gaussian component, consisting of a bright thin disk, with deconvolved FWHM $\simeq$ 360\,pc,  and a fainter thick disk, with deconvolved FWHM $\simeq$ 1.1\,kpc. Statistically significant emission is detected in a region extending up to vertical distances of $\lesssim 1.3-1.4$~kpc at the $3\sigma$ level. \refcom{The thick disk component is well resolved at the current spatial resolution, while the thickness of the thin disk component should be regarded as an upper limit.} 

\item Under the hypothesis that the same CO--to--H$_{\rm 2}$ conversion factor applies to the thin and the thick disks, we estimate that the thick disk can contribute up to $\sim 27\%$ of the total molecular gas mass.

\item The vertical extent of the molecular gas is comparable to that of the \lancom{DIG}, but significantly smaller than that of the neutral atomic hydrogen, which dominates the budget of neutral gas in the halo at larger heights.

\item The molecular thick disk shows no evidence for a significant vertical rotation lag, in contrast to the $\HI$ \red{and $\rm H\alpha$ components}.
The distribution and kinematic properties of the extraplanar CO gas indicate  that this component traces the cumulative effect of multiple gas ejection events driven by \lancom{SF} in the disk over the past few tens of Myr.

\item The spatial association between molecular gas, ionized gas, and dust filaments, together with the limited vertical extent and kinematic properties, strongly favors a stellar-feedback-driven galactic fountain as the dominant mechanism shaping the thick molecular disk. Alternative scenarios involving external accretion, which are required to explain the properties of the $\HI$ gas extending up to $\sim 22$~kpc from the disk, appear less likely for the molecular component.
\end{enumerate}

These results demonstrate that molecular gas can be efficiently lifted into the halo of non-active and non-starburst disk galaxies, where it participates in disk-halo cycling processes. Interferometric observations that do not include a zero-spacing correction would filter out the existence of low-level CO emission coming from putative thick disk components. Future high-sensitivity observations combining single-dish and interferometric data from an additional sample of edge-on galaxies will be crucial in assessing how common thick molecular disks are, and in quantifying their role in galaxy evolution.

\section*{Data availability}
The science-ready data cube and the moment maps shown in Fig.~\ref{fig:moments_co} are fully available through the CDS upon publication of this article

\begin{acknowledgements}
DJL, SGB, MQ, AU and PT acknowledge support from the Spanish grant PID2022-138560NB-I00 funded by MICIU/AEI/10.13039/501100011033/FEDER, EU. \lancom{DJL acknowledges support from the FPI-contract fellowship PREP2022-000479 associated with project PID2022-138560NB-I00.} This work is based on observations carried out with the IRAM 30m telescope. IRAM is supported by INSU/CNRS (France), MPG (Germany) and IGN (Spain). \red{We thank the referee for the helpful feedback and comments that truly helped us to improve the overall quality of the article.}
\end{acknowledgements}

\bibliographystyle{aa}
\bibliography{bibliography}

\begin{appendix}
\onecolumn
\clearpage
\section{Channel maps}

\begin{figure}[!ht]
\centering
    \includegraphics[width=0.85\linewidth]{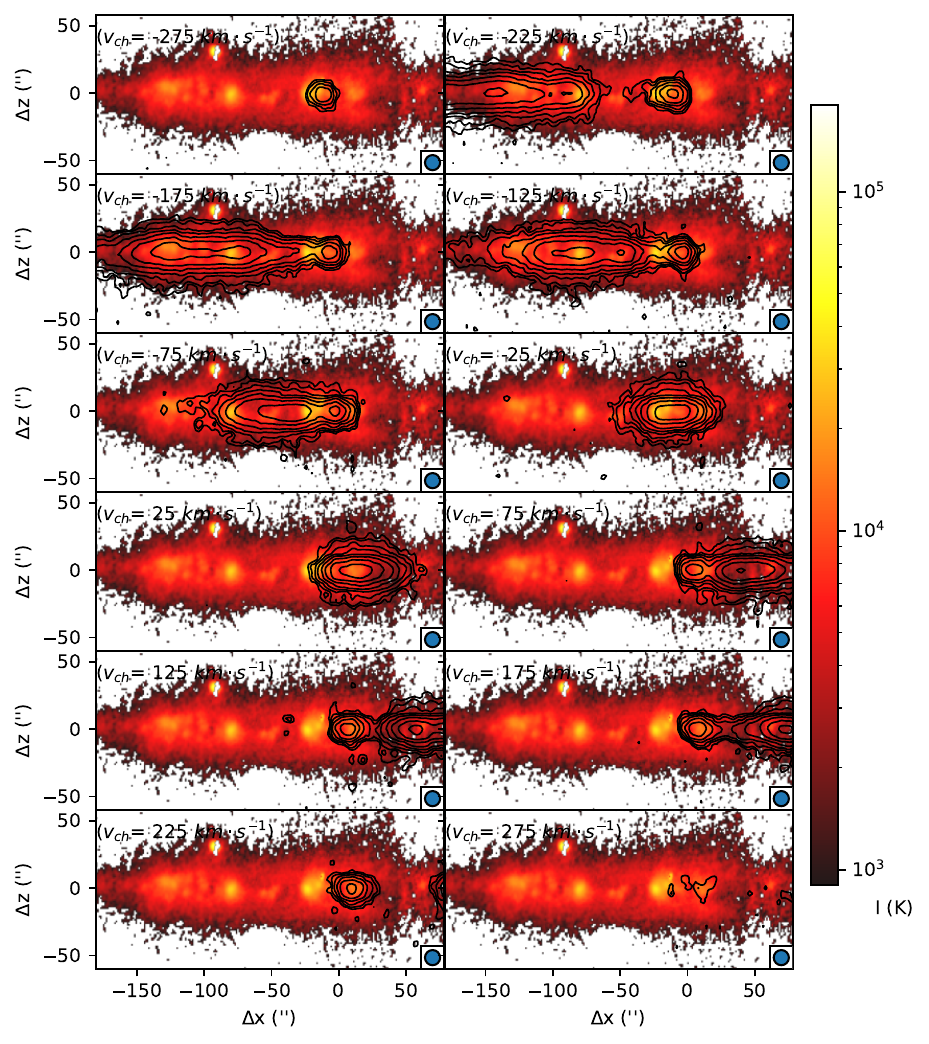}
    \caption{\label{fig:chmap_co_ha}Channel maps of the CO (2-1) intensity of NGC 891, as black contours from 3\,$\sigma_{\rm ch', CO}$ to 600\,$\sigma_{\rm ch', CO}$ in steps of 0.38 dex, where $\sigma_{\rm ch', CO}$ is the \lancom{RMS} noise of the channel maps after a $\Delta$v = 50\,km\,s$^{-1}$ spectral interpolation ($\sigma_{\rm ch', CO}$=3.4$\times$10$^{-3}$ K). The background image shows the H$\alpha$ zeroth-moment map from \cite{Kamp06}.}
\end{figure}

\begin{figure*}[htbp]
\centering
    \includegraphics[width=.85\linewidth]{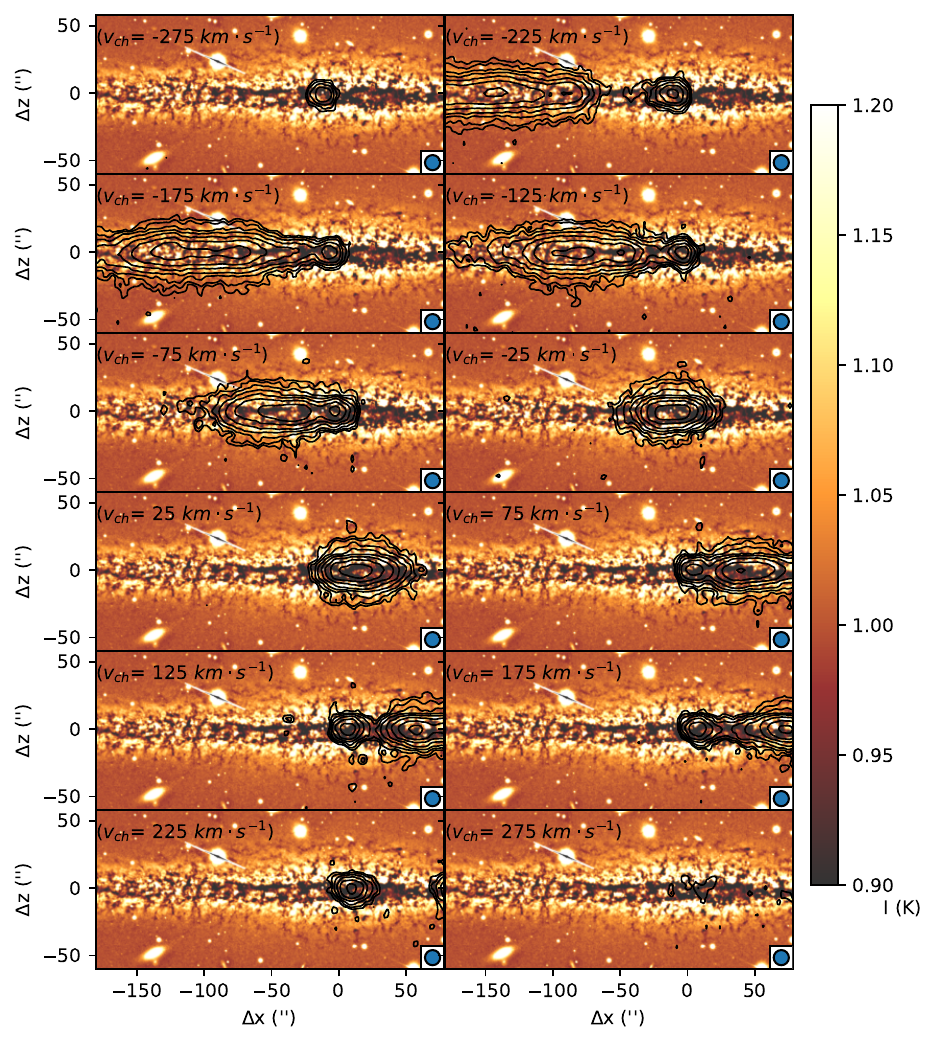}
    \caption{\label{fig:chmap_co_usm}Channel maps of the CO (2-1) intensity of NGC 891 as black contours, the same as the ones on Figure~\ref{fig:chmap_co_ha}. The background image shows the USM of the V-band from \cite{HS00}.}
\end{figure*}

\clearpage
\section{Gaussian fits of the integrated intensity and residuals.}
\begin{figure}[!ht]
\centering
\includegraphics[width=0.85\linewidth]{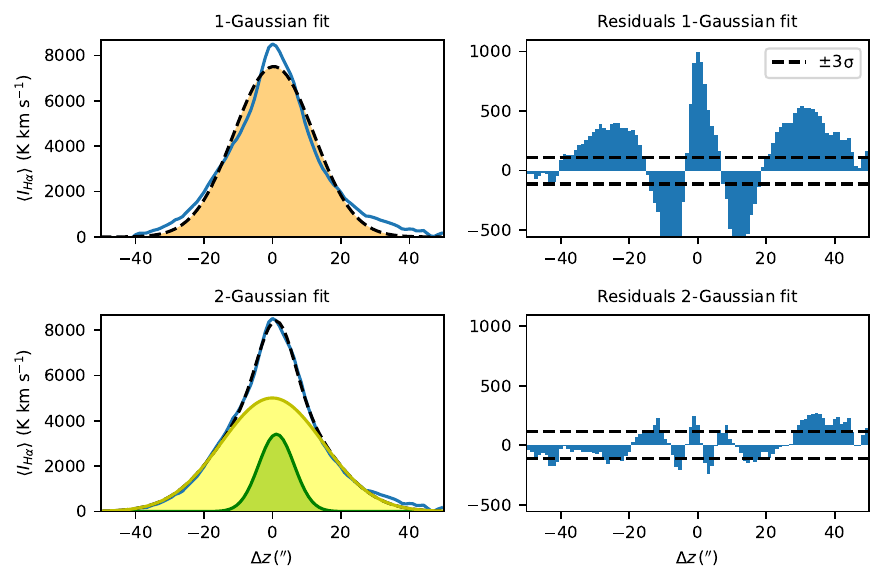}
\includegraphics[width=0.85\linewidth]{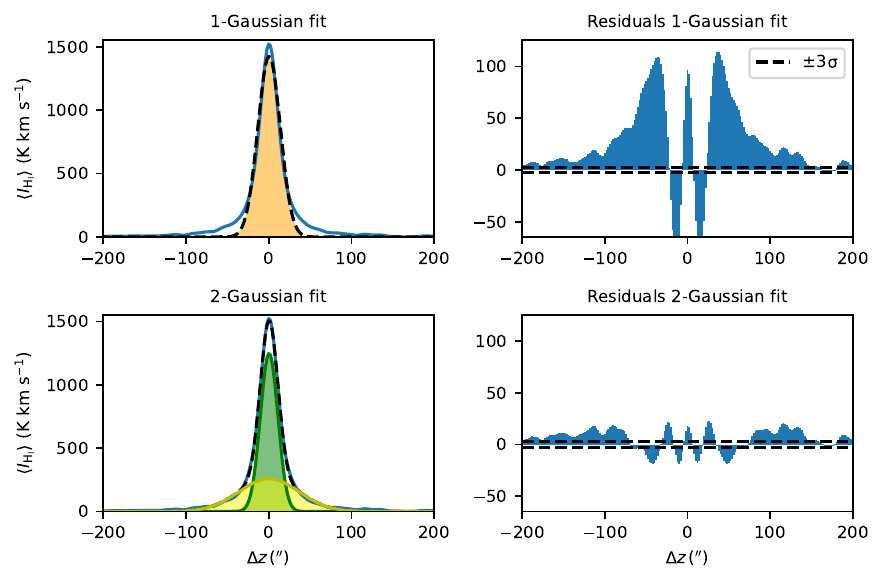}
\caption{\label{fig:gauss_res_ha_hi}Same as Figure~\ref{fig:gauss_res_co} for $\HI$ and H$_\alpha$ ($\sigma_{\rm \HI}$ = 864 mK, $\sigma_{\rm H\alpha}$ = 37 K).}
\end{figure}

\clearpage
\section{PV at different radial slices and velocity spectra at different heights.} \label{multiple-slices}

\begin{figure}[!ht]
\centering  
\noindent%
\includegraphics[width=1\linewidth]{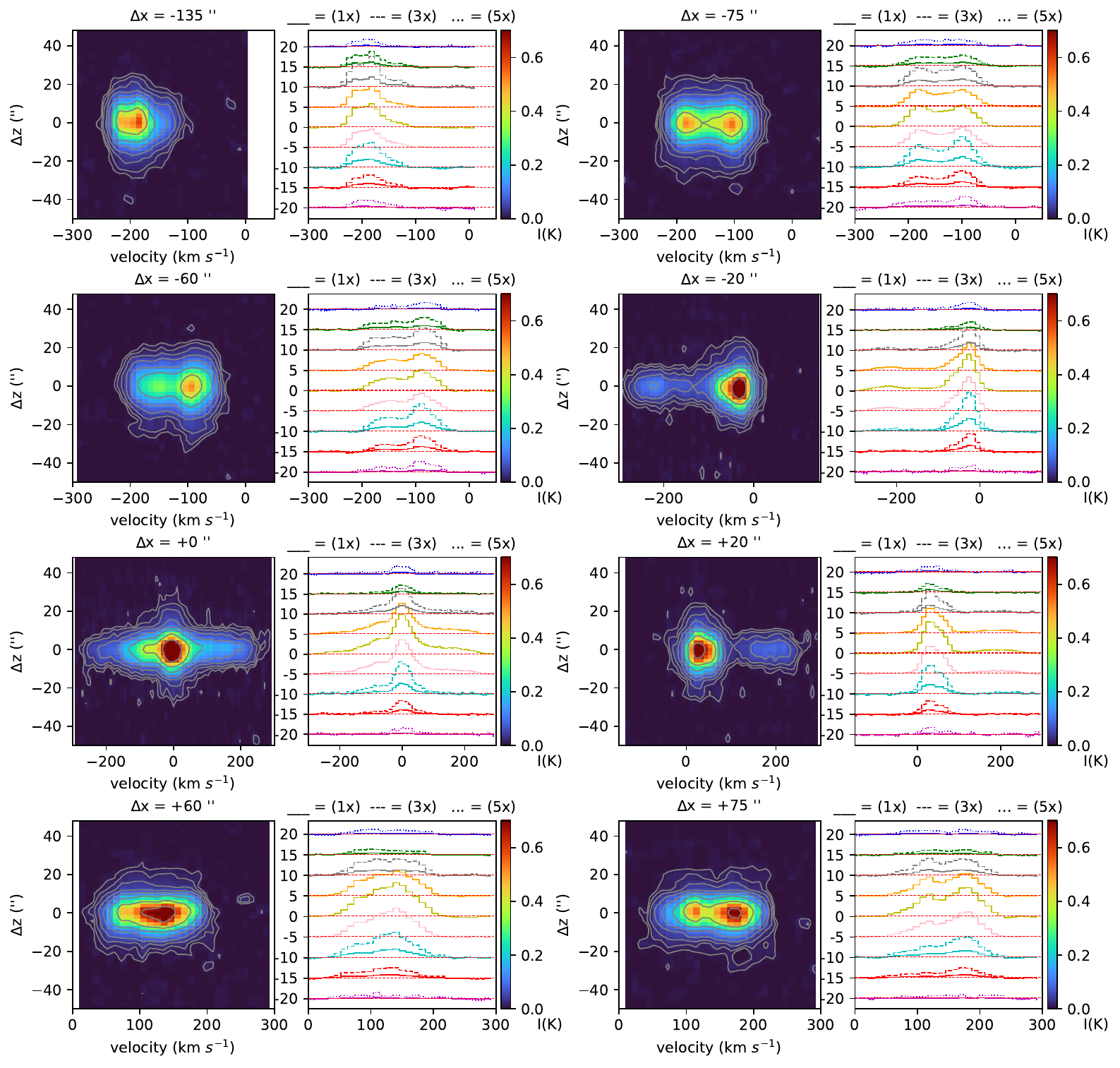}
\caption{\label{fig:pv_spec_CO}Position-velocity slices of CO at different $\Delta x$. \lancom{Gray} contours increase logarithmically from $3\sigma_{\rm ch, CO}$ to $600\sigma_{\rm ch, CO}$ ($\sigma_{\rm ch, CO}$ = 3.5 mK). On the right side of each map: spectra at different $\Delta z$ offsets. Each curve has an arbitrary vertical offset, for clarity. Red dotted horizontal line denotes the $3\sigma_{\rm ch, CO}$ limit of each curve.}
\end{figure}

\begin{figure}[!ht]
\centering  
\noindent%
\includegraphics[width=1\linewidth]{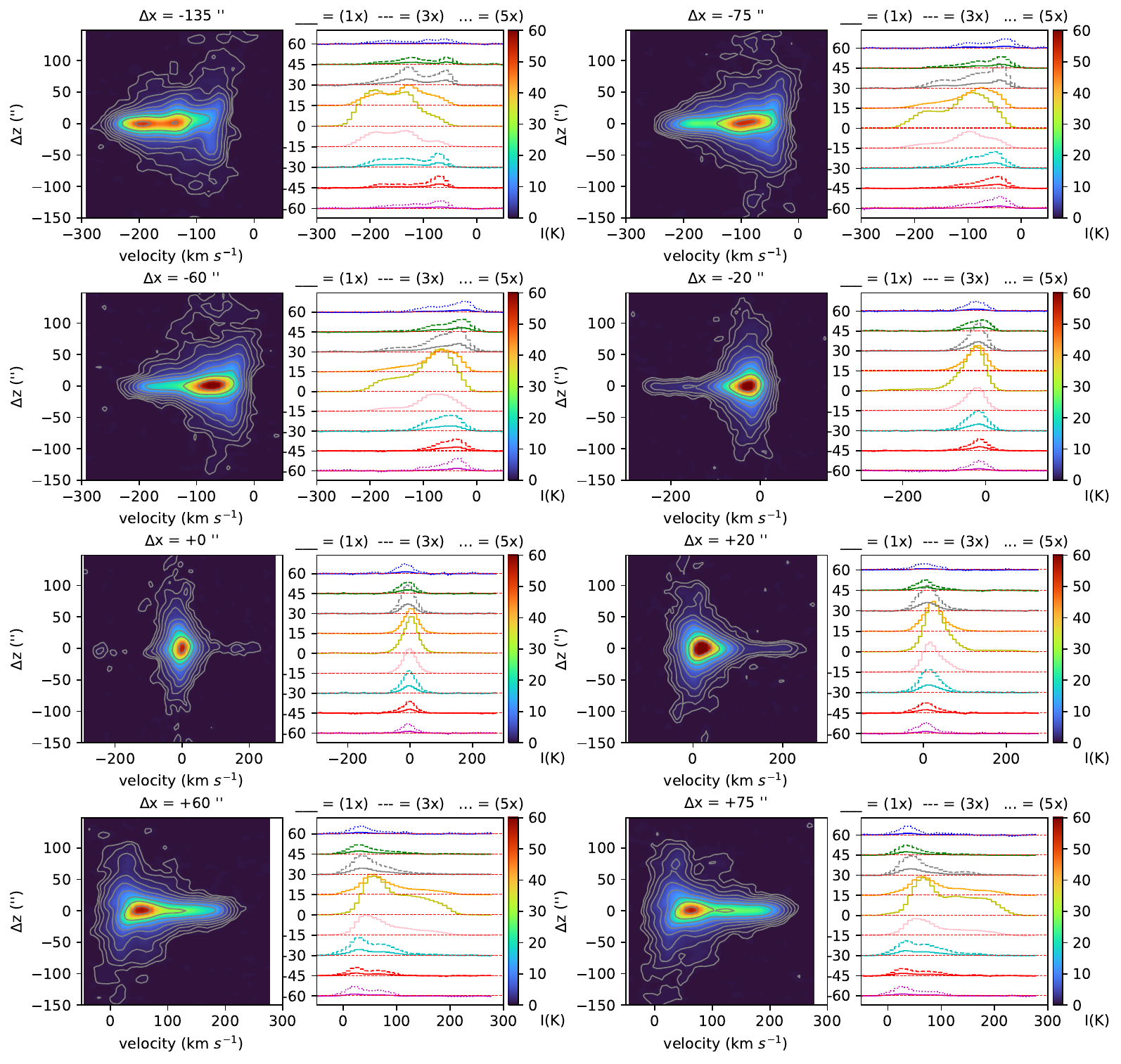}
\caption{\label{fig:pv_spec_HI} Same as \ref{fig:pv_spec_CO} for $\HI$ ($\sigma_{\rm ch, \HI} = 147 mK$).}
\end{figure}

\begin{figure}[!ht]
\centering  
\noindent%
\includegraphics[width=1\linewidth]{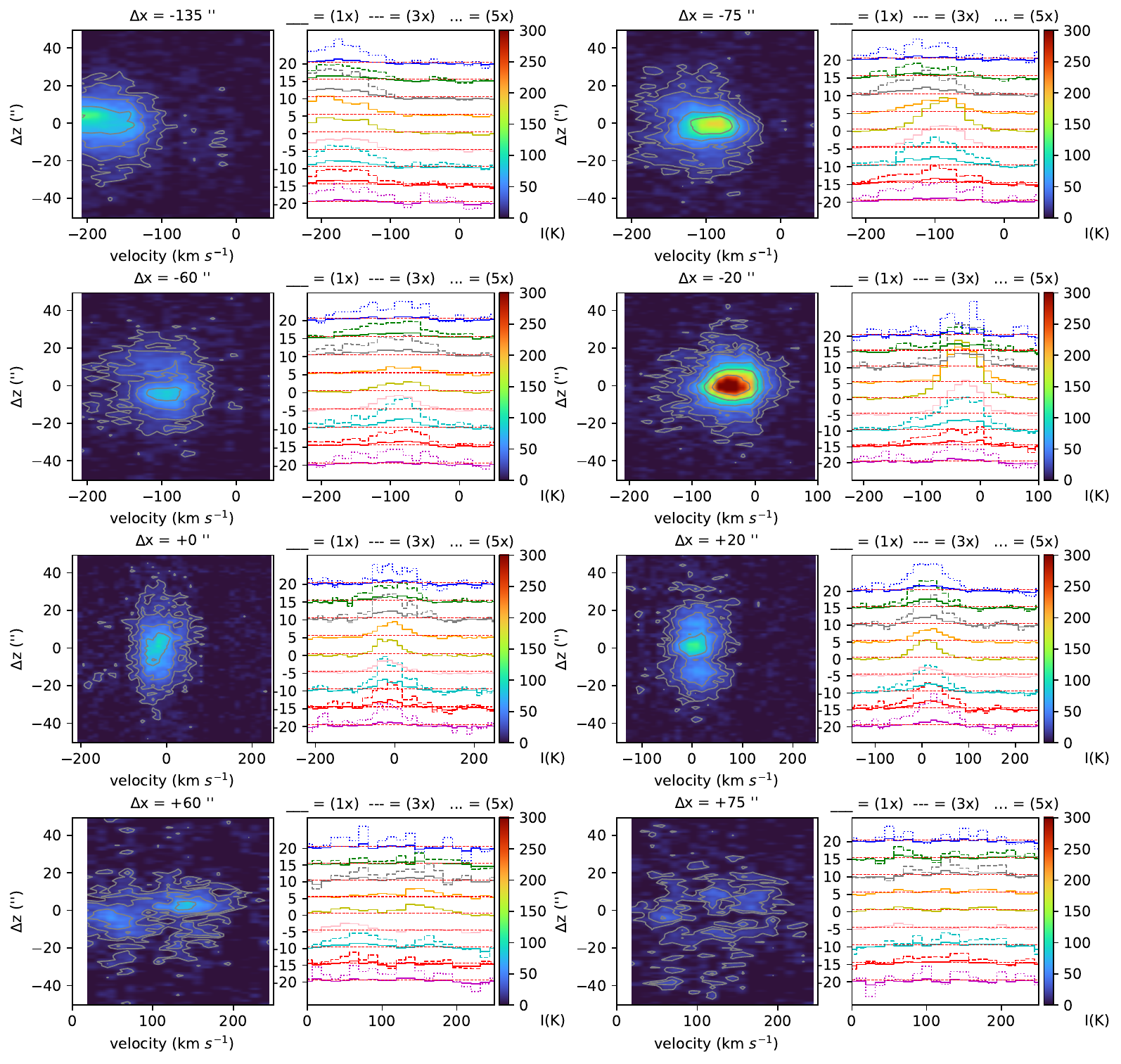}
\caption{\label{fig:pv_spec_Ha}Same as \ref{fig:pv_spec_CO} for H$\alpha$ ($\sigma_{\rm ch, H\alpha} = 3.8 K$).}
\end{figure}

\end{appendix}
\end{document}